\documentclass[fleqn,usenatbib]{mnras}

\usepackage[T1]{fontenc}

\DeclareRobustCommand{\VAN}[3]{#2}
\let\VANthebibliography\thebibliography
\def\thebibliography{\DeclareRobustCommand{\VAN}[3]{##3}\VANthebibliography}

\usepackage{graphicx}
\usepackage{amsmath}
\usepackage{amssymb}

\usepackage{newtxtext,newtxmath}
\usepackage{pifont}
\newcommand{\cmark}{\ding{51}}
\newcommand{\xmark}{\ding{55}}

\newcommand{\edit}[1]{\textcolor{black}{#1}}
\newcommand{\Gaia}{\textit{Gaia}}
\newcommand{\HST}{\textit{HST}}

\title[Astrometric microlensing by LAWD 37]{First semi-empirical test of the white dwarf mass-radius relationship using a single white dwarf via astrometric microlensing}

\author[P. McGill et al.]{Peter McGill$^{1,2}$\thanks{E-mail: pemcgill@ucsc.edu},
Jay Anderson$^{3}$, Stefano Casertano$^{3}$, Kailash C. Sahu$^{3}$,
Pierre Bergeron$^{4}$, \newauthor Simon Blouin$^{5}$, Patrick Dufour$^{4}$, Leigh C. Smith$^{1}$, N. Wyn Evans$^{1}$, Vasily Belokurov$^{1}$, 
\newauthor Richard L. Smart$^{6}$, Andrea Bellini$^{3}$, Annalisa Calamida$^{3}$, Martin Dominik$^{7}$, Noé Kains$^{3}$, Jonas Klüter$^{8}$, \newauthor Martin Bo Nielsen$^{9,10,11}$, and Joachim Wambsganss$^{12}$
\\
$^{1}$Institute of Astronomy, University of Cambridge, Madingley Rd, Cambridge CB3 0HA, UK \\
$^{2}$Department of Astronomy and Astrophysics, University of California, Santa Cruz, CA 95064, USA\\
$^{3}$Space Telescope Science Institute, 3700 San Martin Dr., Baltimore, MD 21218, USA\\
$^{4}$Département de Physique, Université de Montréal, C.P. 6128, Succ. Centre-Ville, Montréal, Québec H3C 3J7, Canada\\
$^{5}$Department of Physics and Astronomy, University of Victoria, Victoria, BC V8W 2Y2, Canada\\
$^{6}$INAF – Osservatorio Astrofisico di Torino, Via Osservatorio 20, 10025 Pino Torinese (TO), Italy\\
$^{7}$University of St Andrews, Centre for Exoplanet Science, SUPA School of Physics \& Astronomy, North Haugh, St Andrews, KY16 9SS, UK\\\
$^{8}$Department of Physics and Astronomy, Louisiana State University, 202 Nicholson Hall, Baton Rouge, LA 70803, USA\\
$^{9}$School of Physics and Astronomy, University of Birmingham, Birmingham B15 2TT, UK\\
$^{10}$Stellar Astrophysics Centre (SAC), Department of Physics and Astronomy, Aarhus University, Ny Munkegade 120, DK-8000 Aarhus C, Denmark\\
$^{11}$Center for Space Science, NYUAD Institute, New York University Abu Dhabi, PO Box 129188, Abu Dhabi, United Arab Emirates\\
$^{12}$Astronomisches Rechen-Institut (ARI), Zentrum fuer Astronomie der Universitaet Heidelberg (ZAH), 69120 Heidelberg, Germany
}

\date{Accepted XXX. Received YYY; in original form ZZZ}
\pubyear{2022}

\begin{document}
\label{firstpage}
\pagerange{\pageref{firstpage}--\pageref{lastpage}}
\maketitle

\begin{abstract}
In November 2019, the nearby single, isolated DQ-type white dwarf LAWD 37 (WD 1142-645) aligned closely with a distant background source and caused an astrometric microlensing event. Leveraging astrometry from \Gaia{} and followup data from the \textit{Hubble Space Telescope} we measure the astrometric deflection of the background source and obtain a gravitational mass for LAWD~37. The main challenge of this analysis is in extracting the lensing signal of the faint background source whilst it is buried in the wings of LAWD~37's point spread function. Removal of LAWD 37's point spread function induces a significant amount of correlated noise which we find can mimic the astrometric lensing signal. We find a deflection model including correlated noise caused by the removal of LAWD~37's point spread function best explains the data and yields a mass for LAWD 37 of $0.56\pm0.08 M_{\odot}$. This mass is in agreement with the theoretical mass-radius relationship and cooling tracks expected for CO core white dwarfs. Furthermore, the mass is consistent with no or trace amounts of hydrogen that is expected for objects with helium-rich atmospheres like LAWD 37. We conclude that further astrometric followup data on the source is likely to improve the inference on LAWD 37's mass at the $\approx3$ percent level and definitively rule out purely correlated noise explanations of the data. This work provides the first semi-empirical  test  of  the  white  dwarf  mass-radius relationship using a single, isolated white dwarf and supports current model atmospheres of DQ white dwarfs and white dwarf evolutionary theory. 
\end{abstract}

\begin{keywords}
gravitational lensing: micro -- astrometry -- white dwarfs
\end{keywords}

\section{Introduction}

Carbon-Oxygen (CO) core white dwarfs are the final evolutionary stage for the vast majority of stars ($\leq 8M_{\odot}$). They are expected to consist mainly of an electron-degenerate core, surrounded by a thin envelope of non-degenerate hydrogen and helium \citep[e.g.,][]{Tremblay2008}. Their mainly degenerate nature means they are expected to follow a mass-radius relationship (MRR) as they evolve and cool. The white-dwarf MRR is important to many areas of astrophysics. It is typically relied upon to calculate the mass of white dwarfs from photometric or spectroscopic measurements \citep[e.g.,][]{Falcon2010} and the white dwarf upper mass limit ($\approx1.4M_{\odot}$) underpins our understanding of the progenitors of type Ia supernovae. Moreover, it is vital when using cooling white dwarfs to date stellar populations in globular clusters \citep[e.g.,][]{Hansen2002}.

To a first-order approximation, in a white dwarf, the inward force of gravity is balanced by the outward pressure of the electron-degenerate gas, resulting in a white dwarf's radius being inversely proportional to the cube root of its mass \citep{Chandrasekhar1935}. Today, detailed evolutionary cooling models including specific degenerate core compositions, the mass of the non-degenerate interior hydrogen layers, and the effects of finite temperature are used to calculate theoretical MRRs \citep[e.g.,][]{Bedard2020}. Despite their sophistication, theoretical MRRs are forced to rely on assumptions about the interior structure of the white dwarf. This is because the masses of the gravitationally stratified non-degenerate interior hydrogen, helium, and CO layers are poorly constrained by observations of the white dwarf's surface. This is particularly important in the case of the mass fraction of the interior hydrogen layer ($q_{\text{H}}$). Depending on temperature and whether a 'thick' or 'thin' hydrogen layer is assumed, theoretical MRRs can vary by $1$-$15$ percent \citep{Tremblay2017}.

For hydrogen-rich white dwarfs (DA type), a thick hydrogen layer is assumed ($q_{\text{H}}=10^{-4}$) which is the estimated maximum hydrogen mass that post-asymptotic-giant-branch evolution models predict for a $0.6M_{\odot}$ white dwarf after residual nuclear burning \citep{Iben1984}. Helium-rich white dwarfs (non-DA type) are either created with hydrogen deficient atmospheres or their hydrogen is hidden beneath their surface \citep{Tremblay2008}. For these white dwarfs, a thin hydrogen layer is assumed ($q_{\text{H}}=10^{-10}$) and represents only trace amounts of hydrogen.

\begin{figure}
    \centering
    \includegraphics[width=\columnwidth]{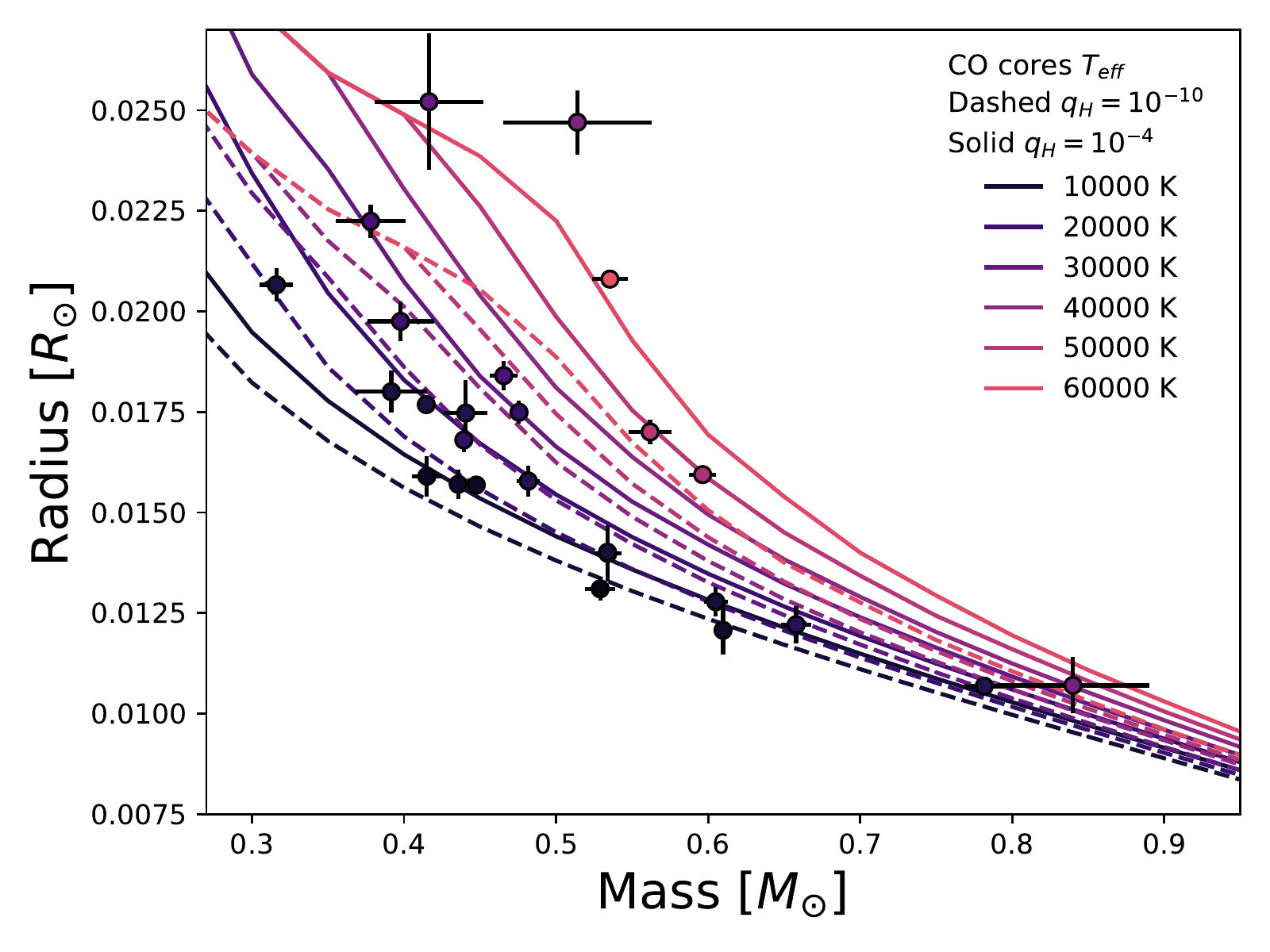}
    \includegraphics[width=\columnwidth]{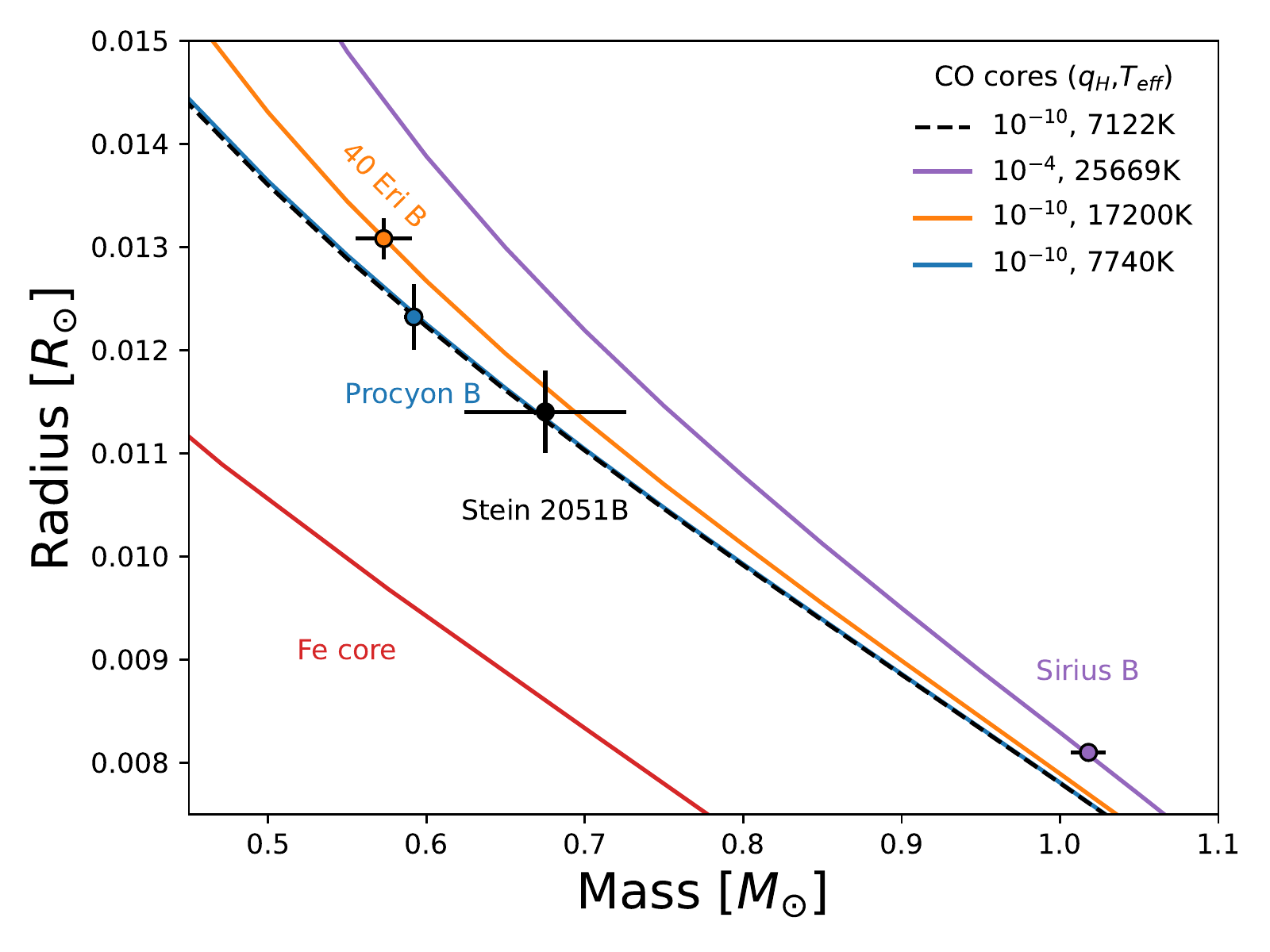}
    \caption[White dwarf mass-radius relationships]{\textbf{Top}: MMRs for 26 white dwarfs in eclipsing binary systems from \cite{Parsons2017} and references therein. Both the masses and radii of these object were determined directly. Figure reproduced from Fig. 9 in \cite{Parsons2017}. \textbf{Bottom}: MRRs for nearby white dwarfs in visual binary systems. The masses for $40$ Eri B, Procyon B, and Sirius B were determined from astrometric measurement of their orbits \citep{Bond2015,Bond2017a, Bond2017b}. The mass for Stein $2051$B was obtained via astrometric microlensing \citep{Sahu2017}. For comparison, the red curve shows the theoretical MRR for zero-temperature white dwarfs with an iron (Fe) core \citep{Hamada1961}. Figure reproduced from Fig. 4 in \cite{Bond2017b}. In both panels, the theoretical MRRs for CO core white dwarfs were obtained from the cooling models of \cite{Bedard2020}.}
    \label{fig:all_mrr}
\end{figure}

Despite the white dwarf MRR's importance, observational tests of the relationship are challenging. The state-of-the-art in direct tests of the white dwarf MMR come from white dwarfs spanning a range of masses and radii that are in eclipsing binary systems \citep[see e.g.,][upper panel in Fig. \ref{fig:all_mrr}]{Parsons2016, Parsons2017}, In these scenarios, both the white dwarf's mass and radius can be determined independently of atmospheric models.\footnote{These MRR tests still require limb darkening coefficients, but the adopted values of the limb darkening coefficients impact the final physical parameters typically below their statistical uncertainties \citep{Parsons2017}} Otherwise, for white dwarfs not in eclipsing binary systems, semi-empirical tests of the MRR which depend on white dwarf atmospheric models, are possible.  

By fitting atmospheric models to broad-band photometry and spectroscopy \citep[e.g.,][]{Giammichele2012}, a white dwarfs' atmospheric parameters ($T_{\text{eff}}$, $\log g$), and its solid angle can be measured. Combining the solid angle with distance information (from parallax), allows the photometric radius of the white dwarf to be measured \citep[e.g.,][]{Kilic2020}. The photometric radius can then be combined with $\log g$ to infer the mass of the white dwarf, and the MRR can be tested \citep{Schmidt1996, Bedard2017, Bergeron2019}. The problem with this approach, however, is that both the mass and the radius are entirely derived from the atmospheric models and it is often difficult to disentangle observed features of the MRR from systematic effects and degeneracies in the atmospheric models \citep{Tremblay2017}. For more robust semi-empirical tests of the MRR, the photometric radius needs to be combined with a mass determination independent of atmospheric models. 

For white dwarfs, it is also possible to obtain MRR information from gravitational redshift measurements \citep{Einstein1916}. The difficulty with this technique is that, for white dwarfs, the shift in the absorption lines due to gravitational redshift is of a similar size to, and degenerate with, the Doppler shift caused by radial motion. This means that the observed shift in spectral lines is a combination of both effects and the gravitational redshift signal can only be isolated if the radial velocity of the white dwarf is precisely known. Determining the radial velocity of a white dwarf is possible when it is in a binary system by taking measurements of its companion \citep[e.g.,][]{Joyce2018}. The gravitational redshift can also be measured for groups of white dwarfs that are co-moving \cite[e.g.,][]{Pasquini2019}, or by averaging over random radial motions \citep[e.g.,][]{Falcon2010,Chandra2020}. However, in all of these cases, a test of the MRR using gravitational redshifts measurements requires an atmospheric model-dependant photometric radius determination.

The most precise direct masses (and semi-empirical tests of the MRR) for white dwarfs that are not post-common envelope (i.e., white dwarfs not in an eclipsing binary system), come from astrometric measurements of visual binary systems. In these particular semi-empirical MRR tests, the radius of the white dwarf is derived by from fitting photometry and spectroscopy and so these tests remain model-dependant and are not entirely from free systematic effects in white dwarf atmospheric models \citep{Tremblay2017}. Fig.~\ref{fig:all_mrr} shows the MRR for nearby white dwarfs in visual binaries with direct mass determinations; $40$ Eri B \citep{Bond2015}, Procyon B \citep{Bond2017a}, and Sirius B \citep{Bond2017b}. Fig.~\ref{fig:all_mrr} also shows Stein $2051$ B, which also happens to be in a visual binary system but its mass has been determined by astrometric microlensing \citep{Sahu2017}. All of these objects have photometric radius determinations, and are in agreement with the theoretical MRRs \citep{Bedard2020}. A semi-empirical test of the MRR for a single and isolated white dwarf has yet to be performed.

Astrometric microlensing events offer unique opportunities to measure the mass of isolated objects \citep[e.g.,][]{Miralda-Escude1996, Kains2017, Rybicki2018, Dong2019, Sahu2022, Lam2022, Kaczmarek2022}. When a foreground lens with mass $M_{\text{L}}$ aligns closely enough with a more distant background source, the  gravitational field of the lens deflects the light of the background source forming a major and minor image of the source. The major image is formed on the same side of the lens as the source and the minor image on the opposite side of the lens. The two images, lens, and unlensed source position always lie along the same line \citep[see e.g.,][for a review]{Bramich2018}. As the lens intervenes between the source and observer, the images change position causing an apparent excursion of the source position \citep[astrometric microlensing;][]{Hog1995, Miyamoto1995, Walker1995}. For the astrometric microlensing event considered in this work, we are in a regime where the major image is resolved from the lens for the duration of the event. In this case the astrometric shift due to microlensing from the unlensed source position is \citep[e.g.,][]{Sahu2017, Bramich2018},
\begin{equation}
    \boldsymbol{\delta}_{+}(\boldsymbol{u}) = \frac{1}{2}\left[\sqrt{u^{2}+4}-u\right]\Theta_{\mathrm{E}}\hat{\boldsymbol{u}}
    \label{eq:resolved_shift}
\end{equation}
Here, $\Theta_{\mathrm{E}}=\sqrt{4Gc^{-2}M_{L}(D^{-1}_{L}-D^{-1}_{S})}$ is the angular Einstein radius of the lensing system \citep{Chwolson1924, Einstein1936}, and $G$, $c$, $D_{L}$, and $D_{S}$ are the gravitational constant, the speed of light, the distance to the lens and the distance to the source, respectively.  $\boldsymbol{u}$ is the angular lens-source separation vector pointing towards the source position in units of $\Theta_{\mathrm{E}}$ or $\boldsymbol{u}=(\boldsymbol{\Phi}_{S}-\boldsymbol{\Phi}_{L})/\Theta_{\mathrm{E}}=\boldsymbol{\beta}/\Theta_{\mathrm{E}}$, where $\Phi_{L,S}$ denotes the angular position of the lens and source, respectively. $u = |\boldsymbol{u}|$ and $\hat{\boldsymbol{u}}$ denotes the unit vector. Crucially, if $\boldsymbol{\delta}_{+}$ can be measured during the event and $D_{L}$ and $D_{S}$ are known, then $M_{\text{L}}$ can be inferred. 

For an astrometric microlensing event to be monitored it first must be found. \cite{Refdal1964} first noted that if the positions and proper motions of celestial objects were known with sufficient accuracy, then close alignments between them, and hence microlensing events, could be predicted ahead of time. This is in contrast to the currently dominant channel of finding photometric microlensing events, in which hundreds of millions of stars are monitored in the Galactic bulge and plane to catch event as they unfold \citep[e.g.,][]{OGLEIV2019,KMTnet2016, Husseiniov2021}.  Following the suggestion of \cite{Refdal1964} and \cite{Paczynski1995, Paczynski1996HST} many attempts to predict microlensing events followed \citep{Feibelman1966, Feibelman1986, Sahu1998, Salim2000, Proft2011, Sahu2014, Proft2016, Lepine2012, Harding2018}. Follow-up of these predictions has proven difficult because of imprecise astrometry which lead to low confidence event predictions. Interest in predicting events was reignited with the advent of astrometry from the \Gaia{} satellite \citep{Prusti2016}, orders of magnitude more precise and numerous than any of its predecessors. First, using \Gaia{} Data Release 1 \citep{GDR12016}, \cite{McGill2018} made one high confidence prediction of an astrometric microlensing event by a nearby white dwarf. Then, following \Gaia{} Data Release 2 \citep{GDR22018} many independent studies searched for both astrometric and photometric events resulting in $\sim5000$ predictions \citep{Kluter2018a, Kluter2018b, Bramich2018, Bramich2018b, Mustill2018, Nielsen2018, Ofek2018, McGill2019a, McGill2019b, McGill2020}. Most recently \cite{Kluter2021} searched for predicted microlensing events using \Gaia{} Early Data Release 3 \citep[GEDR3;][]{GEDR32021} \edit{finding} $1758$ new events\edit{, and \cite{Luberto2022} searched for events with brown dwarf lenses.}

Only two predicted astrometric microlensing events have been successfully followed before this paper. Using the \textit{Hubble Space Telescope} (\HST{}), \cite{Sahu2017} successfully detected the astrometric signal of the microlensing event originally predicted by \cite{Proft2011} involving the nearby white dwarf Stein $2051$B. The astrometric microlensing signal permitted \cite{Sahu2017} to measure a gravitational mass for Stein $2051$ B of $0.675\pm0.051M_{\odot}$. This marked the first ever detection of the astrometric microlensing effect outside the solar system and provided a direct test of white-dwarf evolutionary theory.  Next, \cite{Zurlo2018} detected an astrometric microlensing event using the \textit{Very Large Telescope}. This event was caused by our nearest stellar neighbour, Proxima Centauri, and was originally predicted by \cite{Sahu2014}. Using the astrometric signal, \cite{Zurlo2018} determined the mass of Proxima Centauri to be $0.150^{+0.062}_{-0.051}M_{\odot}$. This marked the first direct gravitational mass measurement of Proxima Centauri, and provided the only current opportunity for a direct mass determination.

In this paper we present analysis of the follow-up of the astrometric microlensing event caused by the nearby DQ-type white dwarf LAWD 37 (\edit{WD 1142-645}), originally predicted by \cite{McGill2018}. This event peaked in November 2019, with a predicted astrometric shift of $\delta_{+}\approx2.8$ mas. First, we describe the two sources of data used in this analysis which are the multi-epoch astrometry from \HST{}, and the astrometric solutions for the source and lens from GEDR3. Next, in contrast to the analysis of the two previous predicted astrometric microlensing in \cite{Sahu2017} and \cite{Zurlo2018}, we detail the combination of these sources of data within a fully Bayesian framework to infer the mass of LAWD 37. Then we describe the concept of leave-one-out cross validation and use it to compare different noise models of the data. We then use the inferred gravitational mass from the astrometric microlensing event as a semi-empirical test of the white dwarf MRR. Finally, we explore the implications of this work on the follow-up of future predicted microlensing events.

\section{Data}\label{sec:data_lawd37}

There are two sources of data used in this study. First, we have multi-epoch HST observations of the lensed source position before, during the predicted maximum of the event, and after.  And, second, we have the astrometric solution of the source and lens from GEDR3. These astrometric solutions provide information on the unlensed source and lens trajectories. The combination of these two sources of data allows the astrometric shift due to microlensing to be measured, and consequently, LAWD~37's mass to be determined. 

\subsection{\HST{} astrometric measurements}
\label{sec:hst_data}

We have nine successful epochs of WFC3/UVIS \HST{} observations spanning over one year. A summary of these observation used to determine the position of the source\footnote{Additionally, a series of WFC3/UVIS F814W short exposures (1.5s) were taken at each epoch in an attempt to constrain the position of the lens which was saturated in the long exposures. Unfortunately there were too few unsaturated, high-signal-to-noise stars common to both exposures needed to constrain the \HST{} focus drift between the two exposure pointings. Therefore, we did not use any \HST{} data to constrain the lens position and instead relied on the projected GEDR3 position of the lens which turned out to be more than adequate for the analysis (see Sections \ref{sec:GEDR3_prior} and \ref{sec:prior_sensitivity}).}
from \HST{} programs GO- 16251, 15961, and 15705 (PI Kailash C. Sahu) is given in Table \ref{tab:hst_obs}.  All \HST{} epochs were taken with moderate-sized dithers ($\pm 100$ pixels) among the pointings.  The dither throw was small enough to allow all exposures to the include about ~20 stars that could serve as astrometric references, but it was large enough to provide some control on Charge Transfer Efficiency (CTE) and distortion residuals.

\begingroup
\setlength{\tabcolsep}{4.5pt}
\begin{table}
    \centering
    \begin{tabular}{l|lllll}
         Epoch & Observation & Filter & Subarray &  Exposure & Number of \\
         &  date & & size (pixels) & time (s) & exposures \\
         \hline
         1 & 1 May 2019  & F814W & 2048x2048 & 95 & 5\\
         2 & 18 Sep 2019 & F814W & 2048x2048  & 82 & 5 \\
         3 & 25 Oct 2019 & F814W & 2048x2048  & 89 & \edit{5} \\
         4 & 10 Nov 2019 & F814W & 2048x2048 & 95 & 5\\
         5 & 26 Nov 2019 & F814W & 1024x1024  & 85 & 13 \\
         6 & 5 Dec 2019  & F814W & 1024x1024 & 85 & 13 \\
         7 & 3 Jan 2020  & F814W & 1024x1024 & 85 & 13 \\
         8 & 5 May 2020  & F814W & 1024x1024 & 85 & 13 \\
         9 & 16 Sep 2020 & F814W & 1024x1024 & 85 & 14 \\
         \hline
    \end{tabular}
    \caption{Summary of the \HST{} WFC3/UVIS observations used to constrain the position of the source. The subarray is the pixel subarray of the UVIS2 detector which were chosen to minimize Charge Transfer Efficiency (CTE) effects. The data are from \HST{} programs GO- 16251, 15961, and 15705 (PI Kailash C. Sahu).}
    \label{tab:hst_obs}
\end{table}
\endgroup

\begin{figure}
    \includegraphics[width=\columnwidth]{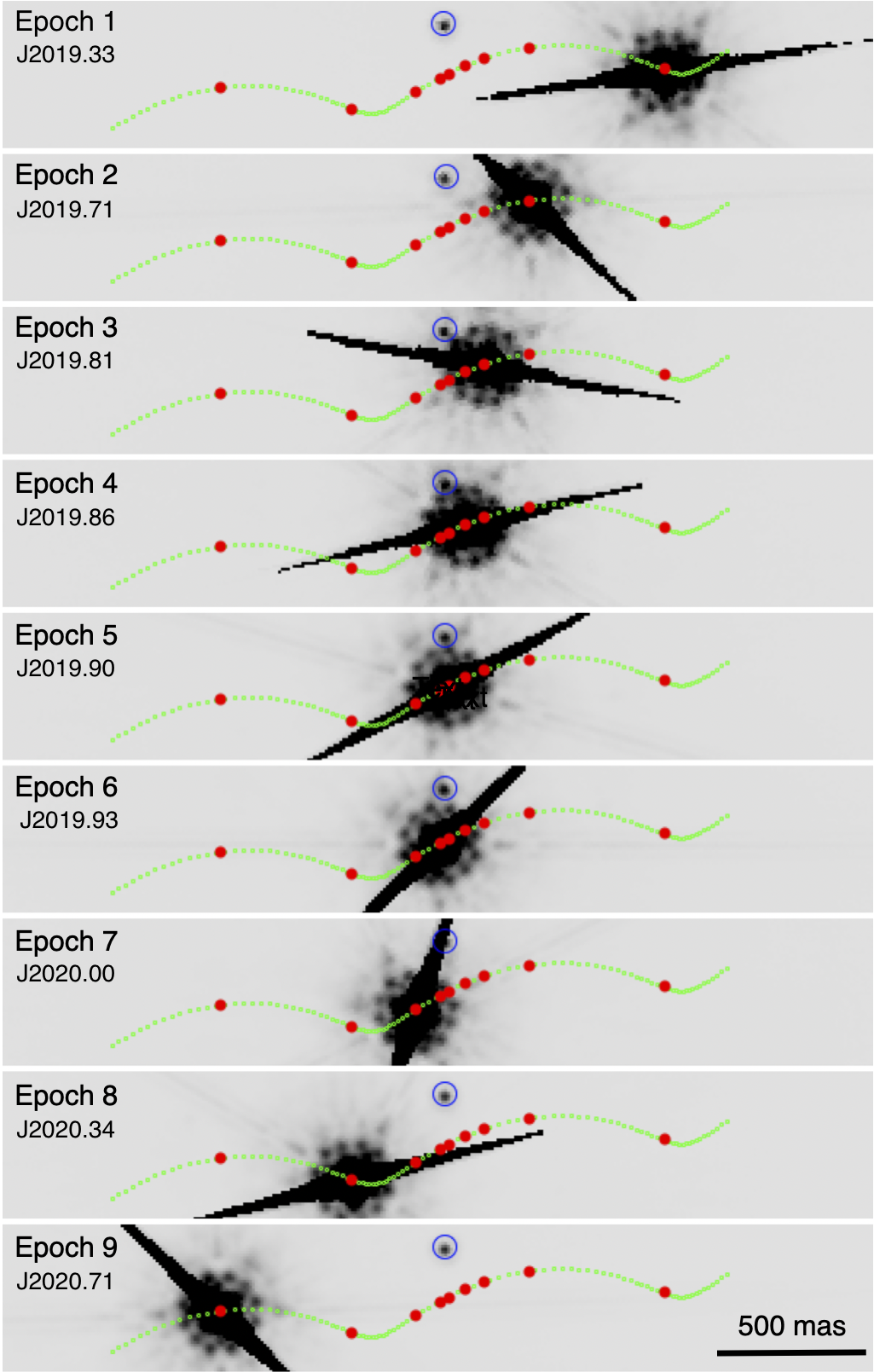}
    \caption[LAWD 37 \HST{} image cutouts]{\HST{} F814W-band image cutouts (co-added stacks by epoch) for each epoch of data during the LAWD 37 event. North is directly upwards in each epoch. The source star is marked with a blue circle. LAWD 37 is the saturated source moving from right to left. The images were made by stacking all images in a given epoch. Dates of each epoch are indicated in Julian years.}
    \label{fig:lawd_74_mosaic}
\end{figure}

Figure~\ref{fig:lawd_74_mosaic} shows the path of LAWD 37 relative to the source star.  In epoch 7, the image of the source is largely occulted by LAWD 37's bleed column.  In epochs 3 through 6, the bump-like features of the point spread function (PSF) introduce considerable complications into our measurement of the source's location.  For this reason, we have developed a model of the extended LAWD 37 PSF in the F814W images.  

\begin{figure*}
    \includegraphics[width=1.0\textwidth]{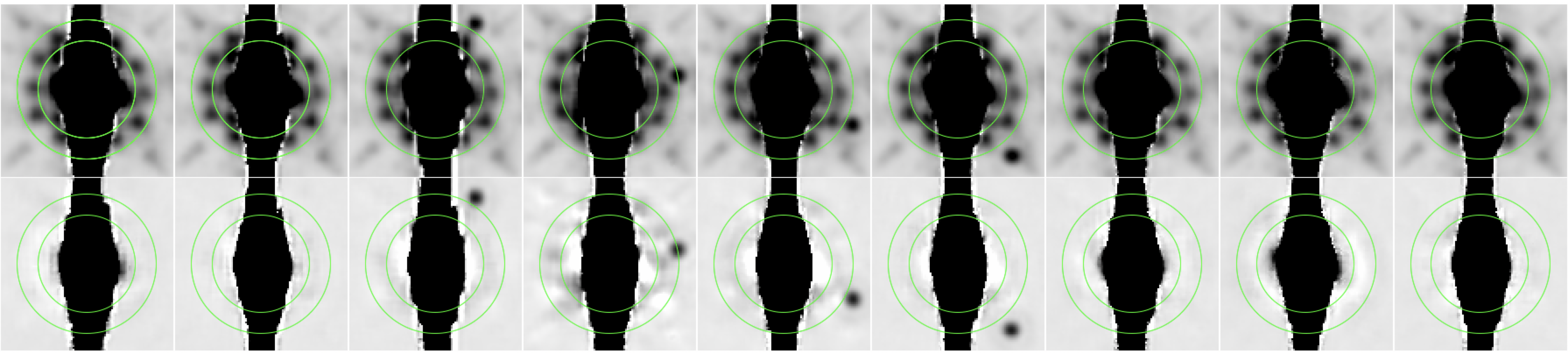}
    \caption[WD PSF]{Top row: Detector-frame cutout imges (co-added stacks by epoch) of the inner ~31$\times$31 pixels (1240$\times$1240 mas) of LAWD 37 for each of the epochs (Epochs 1 to 9 are left to right). The "bumps" in PSF in the annular region between a radius of 7 and 10 pixels (marked in green on the plot) are clear. Bottom row:  The same images, but with the average LAWD 37 PSF subtracted (Epoch 4 was excluded from the average). The source can be seen close to LAWD 37 in the third, fourth, fifth, and sixth epochs. The large vertical black region at the center of each of these images corresponds to the saturated pixels at the center of each deep exposure of LAWD 37.}
    \label{fig:WD_PSF_byE}
\end{figure*}

\edit{We combined together all the F814W images of LAWD 37 from each of the 9 epochs and constructed a 4$\times$-supersampled version of the LAWD 37 image for that epoch. The top row in Fig.~\ref{fig:WD_PSF_byE} shows the 9 stacked images (in the raw-pixel frame, so that the PSF features will all have the same orientation). In order to subtract the PSF of LAWD 37 at each epoch, we used an average of the PSFs constructed from the other 8 epochs (since subtracting the PSF constructed from the  same-epoch images would remove the source, particularly at close separations). As can be seen from Fig. 2, the source positions are well separated from each other in different epochs, so that the position measurements of the source are not  adversely affected by the presence of the source at other epochs. After this subtraction, we noted that most of the subtracted images were quite clean, but the fourth epoch has considerably larger subtraction residuals than the others, presumably due to the telescope experiencing an unusual focus level due to breathing \citep{Anderson2018}.} Unfortunately, this fourth epoch also has the source very close to the lens, which means that the position we measure for the source will be impacted by considerable PSF-subtraction errors. Additionally, the observation in this epoch also suffered from large focus variations. As a result, the PSF subtraction at the position of the source showed large and variable residuals, making the results unreliable. For these reasons, we rejected the fourth epoch from further analysis.

In most of the exposures of epoch $7$, the source's brightest pixels were just offset from the bleed-column and bleed-column-adjacent pixels (see Fig. \ref{fig:lawd_74_mosaic}). Hysteresis in the readout amplifier causes some slight correlation from column to column; this is negligible for pixels of similar brightness, but it can be appreciable for low-value pixels side-adjacent to high-value pixels, such as one finds in bleed columns. When the brightest pixel of a source is available, though, it is possible to measure a position and flux.  Normally when a  position and flux is measured, a 5x5 raster of pixels centered on the star's brightest pixel is used. For epoch $7$, many of these pixels were corrupted by the bleed column, so the fit-box was modified to only include legitimate pixels.

The source can be seen close to LAWD 37 in the third, fourth, fifth, and sixth epochs.  We used the average PSFs from the eight good-focus epochs (with the source and its vicinity masked out in the relevant images) to construct an average F814W PSF for LAWD 37.  Next, this average PSF was subtracted from the LAWD 37 image in each of the individual {\tt \_flc} images, which were corrected for CTE using the pixel-based correction \citep{Anderson2021}. We then fitted the source and the other reasonably bright stars (unsaturated and a signal-to-noise ratio $\gtrsim50$) in each of these images using the F814W "library" PSF described in \cite{Sabbi2013}. The PSF fits were done by a chi-squared-minimization fit of the PSF model\footnote{\url{https://www.stsci.edu/hst/instrumentation/wfc3/data-analysis/psf}} to the star's inner 3$\times$3 pixels, after subtraction of a modal sky taken from an annulus with radii 8 to 12 pixels \citep{Anderson2016}. This small fitting aperture was used in an effort to minimize any influence of the LAWD 37 PSF-subtraction residuals on the fit.  The fitting solved for three parameters:  a flux and a ($x$,$y$) position for each star in the raw frames.  These raw positions were then corrected for distortion \citep{Bellini2011}.

The {\tt \_flc} images were corrected for CTE using the most recent pixel-based correction (v2.0, which is currently in the WFC3/UVIS pipeline).  Even with this correction, there are small residual CTE-related trends present.  Since the various epochs were taken at different telescope orientations, we can inter-compare the measured positions to examine any residual trends. We used linear transformations to map the distortion-corrected positions measured in each exposure in the various epochs into the distortion-corrected frame of the first exposure, using the positions of the unsaturated stars with S/N greater than ~100 in the reference exposure and the individual exposures. We then fitted each star with an average position and proper motion.  We then transformed these modeled positions back into the individual exposures to examine position residuals as a function of raw $y$ position and instrumental magnitude ($m = -2.5\times {\rm log_{10}}F_e$, where $F_e$ is the flux in electrons).  This allowed us to determine a simple table $T$ of relative corrections that removed the residual CTE trends as a function of instrumental magnitude.  This correction had the form $T[m](y_{\rm raw}/2048)$, and its typical amplitude was $\sim$0.01 pixel.

The above analysis allowed us to construct generalized corrections for astrometry in the individual exposures. In order to measure the best possible reference-frame position for the source star over time, we based the final transformations on stars within $\pm 1$ F814W magnitude of the source star's brightness and within 625 pixels (25\arcsec) of the source star's position.  There were 22 such stars.  We used the \Gaia{} catalog to predict a reference-frame location for each of these stars at the epoch of each exposure.  We removed one star that had a predicted parallax greater than 1 mas, just to be sure we are dealing with distant stars.

Using the \Gaia{}-predicted positions in the reference frame and the observed, distortion and CTE-corrected positions of the stars, we solved for a 6-parameter linear transformation from each observed image frame into the reference frame.  We then examined the residuals of this transformation and tweaked the positions and the proper motions of the individual stars to improve the fits.  Note that \HST{} astrometric measurements are more precise than \Gaia{} measurements, particularly for stars at V$\sim$18 or fainter, so it makes sense that the reference-frame positions can be improved via this iteration, while maintaining the average absolute properties of the reference frame (zeropoint, scale, bulk motion, etc) that only \Gaia{} can provide. At this stage, one star was found to have larger than expected residuals, likely due to an unresolved companion.  This star was rejected, which left us with 20 stars to use as the basis for the transformations into the reference frame. These reference stars have a nominal position precision of 0.02 pixel (0.8 mas) in each exposure, due to shot-noise, read-noise, and small errors in the PSF model, all of which are random sources of noise \citep[see Fig. 2 in][]{Bellini2014}. Thus, we have a distortion- and CTE-corrected position for each of 21 stars (the source plus 20 good reference stars) in each of 87 exposure frames.  These positions are then used in the analysis that follows to examine how the source position changes over time.

\begin{figure}
    \centering
    \includegraphics[width=\columnwidth]{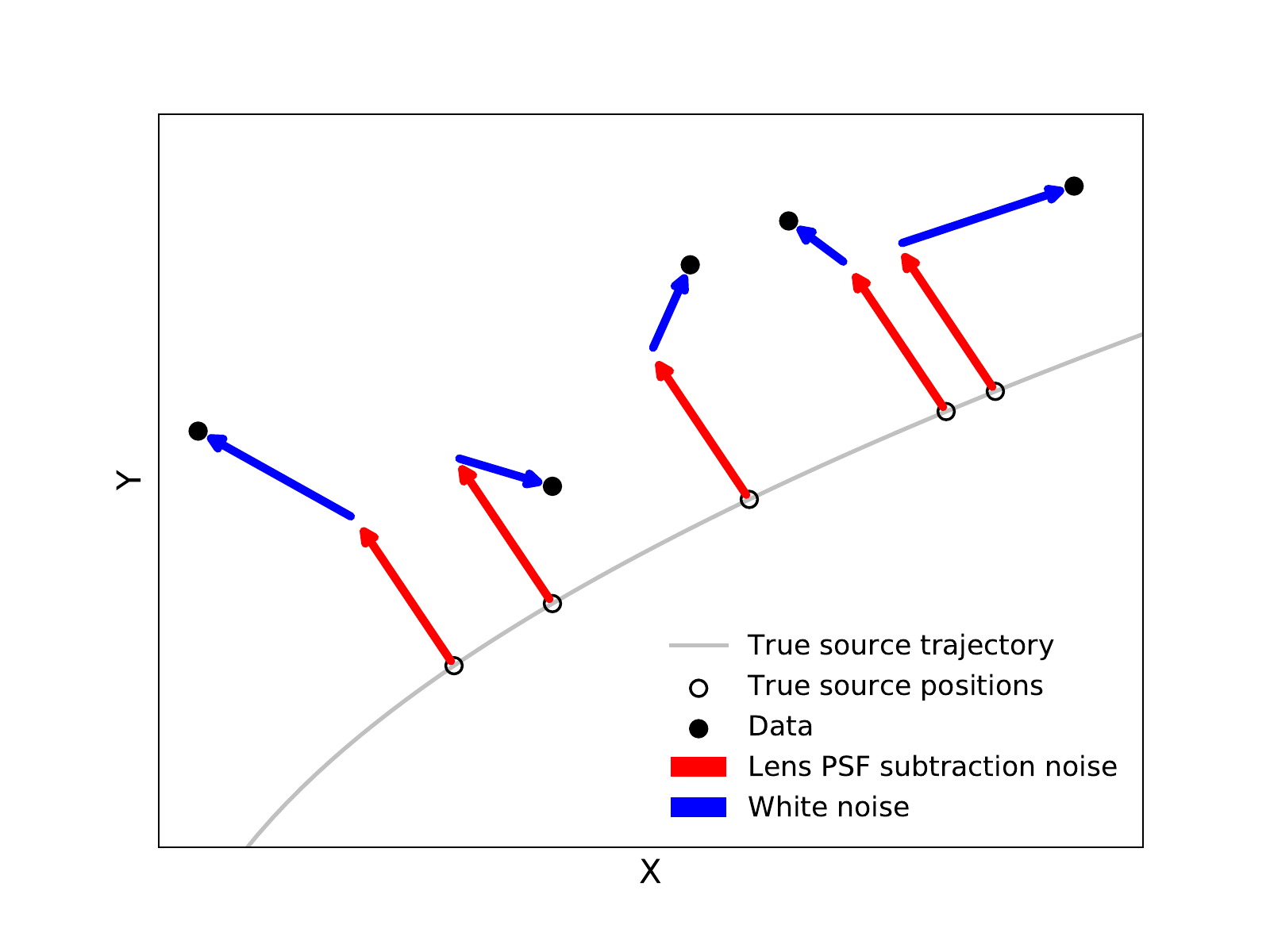}
    \caption[LAWD 37 correlated noise realization]{Schematic of how the source position data are believed to be generated within an epoch. The data receive a correlated error from lens PSF subtraction, which is the same for all data within an epoch (red). The data are also scattered with white noise (blue).}
    \label{fig:noise_realisation}
\end{figure}

Since the PSF is known to vary with time, primarily as a consequence of \HST{} focus drift, there is a variation in the uncertainty in the quality of the PSF for each epoch. It is important to mention that the HST focus does not change gradually and predictably with time. There is an overall trend, but variations during an orbit are larger than the secular variations, and can only be determined after the fact (and with some uncertainty). This means that there are significant PSF variations within an orbit \citep[$\sim6\%$ peak-to-peak maximum variations with respect to the library PSF; see Fig. 5 in][]{Bellini2017}. Critically, this results in a correlated error in the target source position within an epoch. Fig.~\ref{fig:noise_realisation} shows an example realization of noise believed to corrupt the astrometric measurements of the target source within an epoch. The lens PSF subtraction introduces a correlated, within-epoch scatter in addition to the white instrument noise scatter. In order to estimate the size of this correlated error for each epoch, we repeat the PSF subtraction for each epoch with the PSF obtained for each of the other epochs in succession.  We use the distribution of residuals as a proxy for the size of the within-epoch correlated error in the target position. The mean of the distribution of the residuals for each epoch ($m_{\sigma_{e, \text{corr}}}$) are shown in Table \ref{tab:corr_noise}.

In the analysis that follows, all astrometric data are on the tangent plane projected at reference position right ascension $\alpha_{\text{ref}}=176.46045340^{\circ}$, and declination $\delta_{\text{ref}}=-64.84488414^{\circ}$, on the International Celestial Reference System. Coordinate $(\alpha_{\text{ref}},\delta_{\text{ref}})$ corresponds to $(1000,1000)$ pixels in the reference-frame tangent plane with the first coordinate, $X$, having the direction of the local west unit vector and, the second coordinate, $Y$, having the direction of the local north unit vector. Units in both of the coordinates are then scaled to be $1$ mas ($1$pixel=$40$ mas). In what follows we denote a position in this tangent plane as, $\boldsymbol{\zeta}=[X, Y]$.

\begingroup
\setlength{\tabcolsep}{5pt}
\begin{table}
    \centering
    \begin{tabular}{l|llllllll}
         Epoch & $1$ & $2$ & $3$ & $5$ & $6$ & $7$ & $8$ & $9$  \\
         \hline
         $m_{\sigma_{e, \text{corr}}}$ mas & $0.04$ & $0.12$ & $0.6$ & $0.6$ & $0.4$ & $0.6$ & $0.04$ & $0.04$ \\
         \hline
    \end{tabular}
    \caption[PSF correlated noise estimates]{Estimated sizes of the correlated noise standard deviation due to the PSF subtraction in each epoch of data. These values were obtained via simulated PSF subtraction. Note that the size of the correlated noise tends to increase when the lens and source are closest. Epoch $4$ is omitted due to the reasons outlined in Section \ref{sec:hst_data}. The large scatter on epoch $7$ is attributed to the specialized measuring routine used to fit only the uncorrupted pixels due to the source's proximity to a charge bleed column (see Section \ref{sec:hst_data})}
    \label{tab:corr_noise}
\end{table}
\endgroup

\subsection{\Gaia{} astrometric solution}
\label{sec:GEDR3_prior}

GEDR3 provides reference positions, proper motions, and parallax values for both the source and lens. Additionally, each parameter's standard errors and correlations, which are derived from a linear least squares fit of single-epoch astrometric measurements, are provided \citep{LindegrenGEDR32021}. For GEDR3, the astrometric solutions are based on measurements taken between July $2014$ and May $2017$\footnote{\url{https://www.cosmos.esa.int/web/gaia/earlydr3}}. 

%\todo{From Jay:  I'm not sure where the 35,000, etc, coordinates in the vector below come from.  The coordinates for the two objects should be very close to one another, not 10,000 apart in x; not sure what's going on.  The coordinates in the reference frame should be close to (1000,1000); the proper motions and parallaxes should, presumably, be in pixels, not mas.  Maybe there's something I don't undersand, but the text says "tangent plane projection".}

%\todo{From Peter: I scaled everything to be be in units of 1 mas in both the X and Y directions. The position in the vectors below are the reference positions from the EDR3 astrometric solution with reference epoch 2016. Lens has very high proper motion ($>2600$ mas/yr) so it should be pretty far from the projected position and source, at the Gaia reference epoch? ($\approx2600\text{mas/year}\times(2019.9-2016)\text{year}=10140\text{mas}$). All quantities are in mas, not pixels.}

\begin{figure*}
    \includegraphics[width=1.0\textwidth]{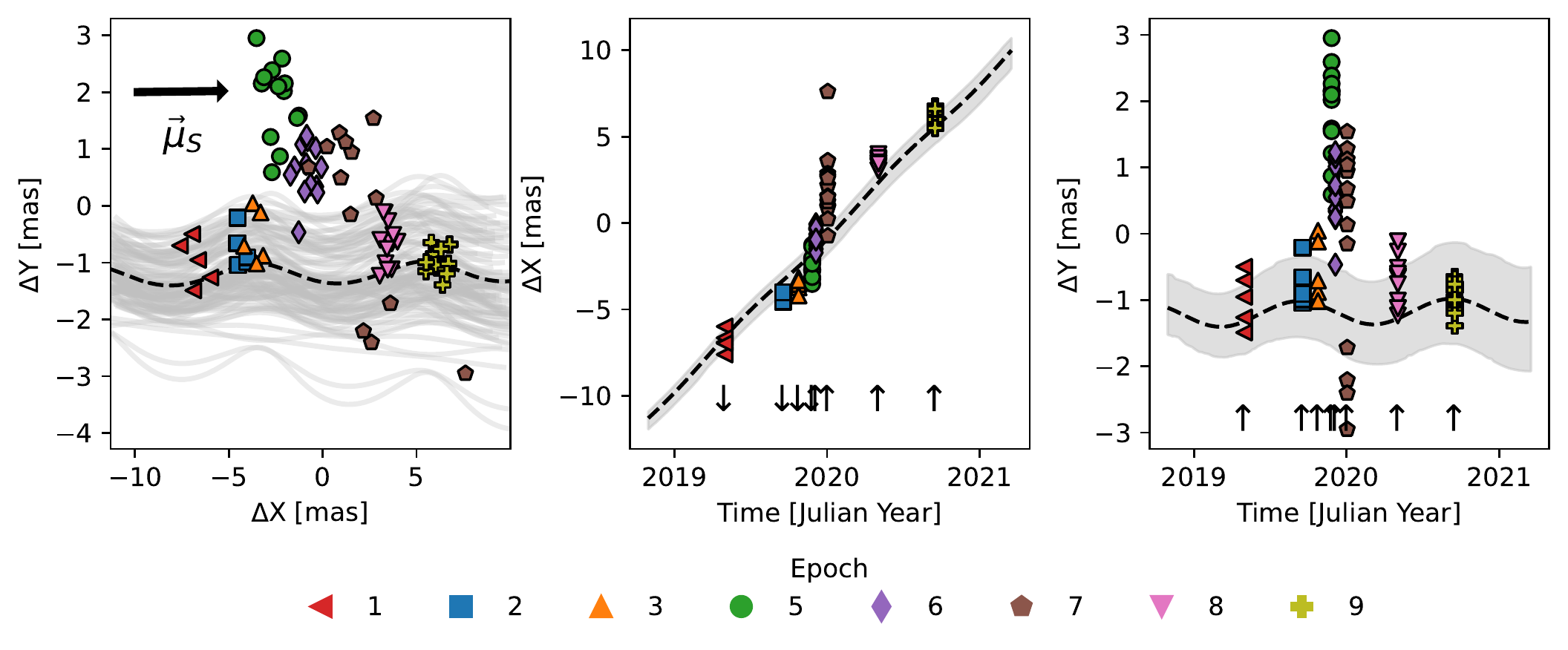}
    \caption[\HST{} follow-up data of the LAWD 37 event]{\HST{} astrometric follow-up data during the predicted microlensing event by LAWD 37. \textbf{Left}: Single \HST{} astrometric measurements coloured by time are shown as circles. Measurements are clustered together in time within eight epochs of data. Also shown are $100$ random samples of the source unlensed projected trajectory from the GEDR3 astrometric solution. Specifically, projected trajectories corresponding to samples $\boldsymbol{\theta}^{\text{ast}}_{S}\sim\mathcal{N}(\boldsymbol{m}^{G}_{S}, \boldsymbol{\Sigma}_{S}^{G})$, are shown in grey and the black dashed line corresponds to $\boldsymbol{\theta}^{\text{ast}}_{S}=\boldsymbol{m}^{G}_{S}$. \textbf{Middle and right}: Projections of the source data in the $X$ and $Y$ directions, respectively. Small arrows indicate the predicted direction of the astrometric deflection signal.}
    \label{fig:data}
\end{figure*}

The mean and covariance matrix for the source (GEDR3 source ID $5332606350796955904$) astrometric parameters in the tangent plane projection defined in Section \ref{sec:hst_data} are,
\begin{align}
    \boldsymbol{m}^{\text{G}}_{S}&=
    \begin{bmatrix}
    35186.9625  \\
    45709.7257 \\
    8.89044 \\ 
    0.03676 \\
    0.19940
    \end{bmatrix}, \\
    \quad{}
    \boldsymbol{\Sigma}^{\text{G}}_{S}&=10^{-1}
    \begin{bmatrix}
     0.2 & -0.01 & 0.04 &-0.06 &-0.04 \\
     -0.01 &  0.2 &-0.05 & 0.003 & 0.05 \\
     0.04 &-0.05 & 0.3 &-0.02 &-0.02 \\
    -0.06 & 0.003 &-0.02 & 0.3 & 0.07 \\
    -0.04 & 0.05 &-0.02 & 0.07 & 0.3 
    \end{bmatrix}.
    \label{eq:source_ast}
\end{align}
Here, the mean and covariance matrices have the parameter order, $X_{0,S}, Y_{0,S}$ source reference positions (mas), $\mu_{X,S}, \mu_{Y,S}$ proper motions in the $X$ and $Y$ directions (mas/year), and $\pi_{S}$ parallax amplitude (mas). Similarly, for the lens (GEDR3 source ID $5332606522595645952$),
\begin{align}
    \boldsymbol{m}^{\text{G}}_{L}&=
    \begin{bmatrix}
    45825.73799 \\
    46592.48286 \\
    -2661.63959 \\  
    -344.93250 \\
    215.67527
    \end{bmatrix},\\
    \quad{}
    \boldsymbol{\Sigma}^{\text{G}}_{L}&=10^{-3}
    \begin{bmatrix}
    0.2   & -0.03 & -0.02 & -0.04 & -0.03 \\
    -0.03 & 0.2   & -0.03 & 0.01  & 0.07 \\
    -0.02 & -0.03 & 0.3   & -0.05 & 0.02 \\
    -0.04 & 0.01  & -0.05 & 0.4   & 0.06 \\
    -0.03 & 0.07  & 0.02  & 0.06  & 0.3 \\
    \end{bmatrix}.
    \label{eq:lens_ast}
\end{align}
This astrometric information from GEDR3 for both the source and lens will be used in Gaussian priors for the models described in Section \ref{sec:models_lawd37}. Fig. \ref{fig:data} shows the GEDR3 predicted unlensed source trajectory, and the \HST{} astrometric measurements of the source. We can see that there are clear offsets from the GEDR3 unlensed source trajectory in the expected direction of predicted lensing signal, although the data are clearly noisy.

\section{Models}\label{sec:models_lawd37}

Fig.~\ref{fig:data} shows that we are in a low signal-to-noise regime, as the offset in the data relative to the projected GEDR3 unlensed position of the source is comparable to the size of the scatter at each epoch. It is therefore important that we investigate a range of models. In this section we describe the different models that were fitted to the data, and compared to one another. We fit models with and without the astrometric lensing signal and with and without a correlated noise component. We first address the choice of parameterisation of the astrometric lensing signal. Next, we outline the different likelihoods used in each of the models. Finally, we detail the prior distributions used in each model and the methods used to sample the posterior distributions in all models.

\subsection{Parameterisation of the microlensing signal}\label{sec:parametisation}

Using the expression for $\boldsymbol{\delta}_{+}$ (Eq.~\ref{eq:resolved_shift}), and calculating $\boldsymbol{\beta}$, using the lens and source trajectories, the lensed source position can be seen to be dependent on $11$ parameters,
\begin{equation}
    \left[X_{0,S}, Y_{0,S}, \mu_{X,S}, \mu_{Y,S},\pi_{S},X_{0,L}, Y_{0,L}, \mu_{X,L}, \mu_{Y,L},\pi_{L}, M_{L}\right].
\end{equation}
Specifically, the source ($\boldsymbol{\theta}^{\text{ast}}_{S} = [X_{0,S}, Y_{0,S}, \mu_{X,S}, \mu_{Y,S},\pi_{S}]$) and lens ($\boldsymbol{\theta}^{\text{ast}}_{L} = [X_{0,L}, Y_{0,L}, \mu_{X,L}, \mu_{Y,L},\pi_{L}]$) astrometric parameters and the lens mass, $M_{\text{L}}$. However, modeling the signal with these parameters is not straightforward. This is due to the fact that parallax enters the model in two different ways. Firstly, the source and lens parallaxes control the lens and source unlensed trajectories and hence the lens-source angular separation. Secondly, the lens and source parallaxes enter as distance terms and control the size of $\Theta_{\mathrm{E}}$. The problem arises due to the interpretation of negative source parallax values when trying to include the GEDR3 astrometric solution as priors in the model. 

The GEDR3 reported value of the source parallax with standard error is $0.20\pm0.16$ mas, where some of the distribution $\pi_{S} < 0$ (assuming a Gaussian distribution). Using negative parallax values when $\pi_{S}$ enters the model in the source trajectory is fine as this reflects uncertainty in the parallax component of source trajectory and is physical. However, a negative $\pi_{S}$ value entering the model as a distance term and re-scaling $\Theta_{\mathrm{E}}$, is not physical. In this case, a negative $\pi_{S}$ value would act to artificially increase $\Theta_{\mathrm{E}}$ and therefore potentially bias the inference towards lower $M_{\text{L}}$. 

There are a number of ways to mitigate this problem. A simple solution is that $\Theta_{\mathrm{E}}$ could be fitted for instead of $M_{\text{L}}$. In this case the astrometric parameters of the model would be $[\boldsymbol{\theta}^{\text{ast}}_{S},\boldsymbol{\theta}^{\text{ast}}_{L}, \Theta_{\mathrm{E}}]$. Here the prior on $\pi_{S}$ only enters the model as a trajectory term and hence negative values are permitted. Finally, $M_{\text{L}}$ can then be extracted after inferring a value for $\Theta_{\mathrm{E}}$.

\subsection{Likelihoods}

In order to model the data, we need to setup a likelihood function that encodes our beliefs as to how the data were generated. Specifically, for all models, we assume the process that generated the data at time $t_{i}$, for the source is of the form,
\begin{equation}
\boldsymbol{\zeta}^{\text{obs}}(t_{i};\boldsymbol{\theta}, \mathcal{M}_{TN}) = \boldsymbol{\zeta}^{\text{T}}(t_{i};\boldsymbol{\theta}^{\text{ast}}) + \boldsymbol{\epsilon}^{N}(t_{i};\boldsymbol{\theta}^{\text{noise}}).
\label{eq:data_generation}
\end{equation}
Here, $\boldsymbol{\zeta}^{\text{obs}}$ is the observed source position in the tangent plane given by model $\mathcal{M}_{TN}$ with trajectory component $T$ and noise component $N$. $\boldsymbol{\zeta}^{\text{T}}$ is the unlensed source position predicted by trajectory component $T$ with astrometric parameters $\boldsymbol{\theta}^{\text{ast}}$. $\boldsymbol{\epsilon}^{N}$ is an additive zero-mean Gaussian noise component $N$ with parameters $\boldsymbol{\theta}^{\text{noise}}$. We consider two different trajectory models, with and without the astrometric lensing deflection term. The model without the deflection is, $\boldsymbol{\zeta}^{\text{T}}(t_{i};\boldsymbol{\theta}^{\text{ast}})=\boldsymbol{\zeta}^{\text{N}}(t_{i};\boldsymbol{\theta}^{\text{ast}})\equiv\boldsymbol{\zeta}(t_{i};\boldsymbol{\theta}^{\text{ast}}_{S})$. Here,
\begin{equation}
    \boldsymbol{\zeta}(t_{i};\boldsymbol{\theta}^{\text{ast}}_{S}) = \begin{bmatrix}
    X_{0,S} \\
    Y_{0,S}
    \end{bmatrix} +
    (t_{i} - t_{\text{ref}})
    \begin{bmatrix}
    \mu_{X, S} \\
    \mu_{Y, S}
    \end{bmatrix} +
    \pi_{S} \textbf{J}^{-1}\boldsymbol{R}_{\oplus}(t_{i})
    \label{eq:tangent_plane_motion}
\end{equation}
$\boldsymbol{R}_{\oplus}(t)$ are Cartesian Barycentric solar system coordinates in au of the
Earth at time $t$. $\boldsymbol{R}_{\oplus}(t)$ was retrieved via the Astropy Python package \citep{AstropyCollaboration2013,AstropyCollaboration2018} which uses values computed from NASA JPL’s Horizons Ephemeris\footnote{\url{https://ssd.jpl.nasa.gov/?horizons}}. $\boldsymbol{J}^{-1}$ is the inverse Jacobian matrix of the transformation from Cartesian to spherical coordinates \citep[e.g.,][Section 7.2.2.3]{Urban2014} projected at the reference position  ($\alpha_{\text{ref}},\delta_{\text{ref}}$), and $t_{\text{ref}}=J2016.0$ is the GEDR3 reference epoch. Eq.~(\ref{eq:tangent_plane_motion}) is just the standard motion of the source with proper motion and parallax. The model with the deflection is  $\boldsymbol{\zeta}^{\text{T}}(t_{i};\boldsymbol{\theta}^{\text{ast}})=\boldsymbol{\zeta}^{\text{D}}(t_{i};\boldsymbol{\theta}^{\text{ast}})\equiv\boldsymbol{\zeta}(t_{i};\boldsymbol{\theta}^{\text{ast}}_{S}) + \boldsymbol{\delta}_{+}(t_{i}; [\boldsymbol{\theta}^{\text{ast}}_{S},\boldsymbol{\theta}^{\text{ast}}_{L},\Theta_{\mathrm{E}}])$ where $  \boldsymbol{\theta}^{\text{ast}}\equiv[\boldsymbol{\theta}^{\text{ast}}_{S},\boldsymbol{\theta}^{\text{ast}}_{L},\Theta_{\mathrm{E}}]$.

For all models, we assume that the noise is uncorrelated between the $X$ and $Y$ directions. With this in mind, we can write the likelihood for a set of $K_{e}$ data points within an epoch, $e$. Let the data in epoch $e$ be denoted as $D_{e} = \{\boldsymbol{t}_{e},\boldsymbol{X}_{e}, \boldsymbol{Y}_{e}\}$, where the elements of $D_{e}$ are vectors of length $K_{e}$, and represent the times, $X$ positions and $Y$ positions of all data points within epoch $e$, respectively. The likelihood of data within an epoch $e$ is then,
\begin{align}
\begin{split}
        p(D_{e}|\boldsymbol{\theta}_{e}, \mathcal{M}_{TN}) =& \mathcal{N}\left(\boldsymbol{X}_{e} |\boldsymbol{X}^{T}_{e}(\boldsymbol{t}_{e};\boldsymbol{\theta}^{\text{ast}}), \boldsymbol{\Sigma}^{N}_{e}(\boldsymbol{\theta}^{\text{noise}}_{e})\right)\\
        &\times\mathcal{N}\left(\boldsymbol{Y}_{e}|\boldsymbol{Y}^{T}_{e}(\boldsymbol{t}_{e};\boldsymbol{\theta}^{\text{ast}}), \boldsymbol{\Sigma}^{N}_{e}(\boldsymbol{\theta}^{\text{noise}}_{e})\right)  
\end{split}
        \label{eq:like_direction}
\end{align}
Here $\mathcal{N}$ is the $K_{e}$ dimensional multivariate Gaussian density, $\boldsymbol{\theta}_{e}\equiv[\boldsymbol{\theta}^{\text{ast}},\boldsymbol{\theta}^{\text{noise}}_{e}]$, $\boldsymbol{X}^{T}_{e}(\boldsymbol{t}_{e};\boldsymbol{\theta}^{\text{ast}})$ and $\boldsymbol{Y}^{T}_{e}(\boldsymbol{t}_{e};\boldsymbol{\theta}^{\text{ast}})$ are vectors of $X$ and $Y$ source positions of length $K_{e}$ obtained by evaluating the vector of times $\boldsymbol{t}_{e}$ for the trajectory model component $T$. $\boldsymbol{\Sigma}^{N}_{e}(\boldsymbol{\theta}^{\text{noise}})$ is a $K_{e}\times K_{e}$ covariance matrix. We consider two noise models for the covariance matrix. An uncorrelated white noise model with $\boldsymbol{\Sigma}^{N}_{e}(\boldsymbol{\theta}^{\text{noise}}_{e})=\boldsymbol{\Sigma}^{W}_{e}(\boldsymbol{\theta}^{\text{noise}}_{e})\equiv\sigma^{2}_{\text{white}}\boldsymbol{I}$, where $\boldsymbol{I}$ is the $K_{e}\times K_{e}$ unit diagonal identity matrix, and $\boldsymbol{\theta}^{\text{noise}}_{e}\equiv[\sigma_{\text{white}}]$. We also consider a correlated and white noise model with $\boldsymbol{\Sigma}^{N}_{e}(\boldsymbol{\theta}^{\text{noise}}_{e})=\boldsymbol{\Sigma}^{C}_{e}(\boldsymbol{\theta}^{\text{noise}}_{e})\equiv\boldsymbol{\Sigma}^{W}_{e}([\sigma_{\text{white}}])+\sigma^{2}_{e,\text{corr}}\boldsymbol{1}$, where $\boldsymbol{1}$ is the $K_{e}\times K_{e}$ full ones matrix, and $\boldsymbol{\theta}^{\text{noise}}_{e}\equiv[\sigma_{\text{white}},\sigma_{e,\text{corr}}]$. The correlated noise model corresponds to the data generation process shown in Fig.~\ref{fig:noise_realisation}.

For all models, under the assumptions of independence between $N_{e}$ epochs of data, the likelihood over the full data set is,
\begin{equation}
    p(\mathcal{D}|\boldsymbol{\theta},\mathcal{M}_{TN}) = \prod_{e=1}^{N_{e}}p(D_{e}|\boldsymbol{\theta}_{e}, \mathcal{M}_{TN}).
    \label{eq:likelihood}
\end{equation}
Here, $\boldsymbol{\theta}$ are all the model parameters and $\mathcal{D}=
\{D_{e}\}_{e=1}^{N_{e}}$ is the full data set over all epochs. The consideration of two deflection model components and two noise model components leads to four distinct models to be investigated: a model with the deflection and correlation noise - $\mathcal{M}_{DC}$; a model with the deflection and just white noise - $\mathcal{M}_{DW}$; a model without the deflection and correlated noise - $\mathcal{M}_{NC}$, and a model without the deflection and just white noise - $\mathcal{M}_{NW}$. Table \ref{tab:models} contains a summary of the different models and their components. 

\begin{table}
    \begin{tabular}{p{1.7cm}|p{1.2cm}p{1.2cm}p{1.2cm}p{1.2cm}}
         &  \multicolumn{4}{c}{Model} \\
         & $\mathcal{M}_{DC}$ & $\mathcal{M}_{DW}$ & $\mathcal{M}_{NC}$ & $\mathcal{M}_{NW}$  \\
         \hline
         Deflection & \cmark & \cmark & \xmark & \xmark \\ 
         White noise & \cmark & \cmark & \cmark & \cmark \\
         Correlated noise & \cmark & \xmark & \cmark & \xmark \\
         \# of parameters & $20$ & $12$ & $14$ & $6$ \\
         parameters $\boldsymbol{\theta}$ & $\boldsymbol{\theta}^{\text{ast}}_{S}$, $\boldsymbol{\theta}^{\text{ast}}_{L}$, $\Theta_{\mathrm{E}}$, $\sigma_{\text{white}}$,  $\boldsymbol{\sigma}_{\text{corr}}$ & $\boldsymbol{\theta}^{\text{ast}}_{S}$, $\boldsymbol{\theta}^{\text{ast}}_{L}$, $\Theta_{\mathrm{E}}$,$\sigma_{\text{white}}$ & $\boldsymbol{\theta}^{\text{ast}}_{S}$, $\sigma_{\text{white}}$, $\boldsymbol{\sigma}_{\text{corr}}$ & $\boldsymbol{\theta}^{\text{ast}}_{S}$, $\sigma_{\text{white}}$\\
        \hline
    \end{tabular}
    
    \caption[LAWD~37 model summaries]{Summary of the components of the four considered models. Deflection indicates if the model contains the astrometric microlensing deflection term. $\boldsymbol{\sigma}_{e}=[\sigma_{1,\text{corr}},\sigma_{2,\text{corr}}, ...,\sigma_{N_{e},\text{corr}}]$ is the vector of correlated noise parameters.}
    \label{tab:models}
\end{table}

\subsection{Priors}

For the source and lens astrometric parameters ($10$ parameters total), there is prior information from GEDR3. Specifically we assume multivariate normal distributions, $p(\boldsymbol{\theta}^{\text{ast}}_{S})= \mathcal{N}(\boldsymbol{m}^{\text{G}}_{S}, \boldsymbol{\Sigma}^{G}_{S}),$ and $p(\boldsymbol{\theta}^{\text{ast}}_{L})= \mathcal{N}(\boldsymbol{m}^{\text{G}}_{L}, \boldsymbol{\Sigma}^{G}_{L})$
using the values in Eqs. (\ref{eq:source_ast}) and (\ref{eq:lens_ast}). 

There are three potential issues that need to be considered when using the GEDR3 astrometric solution as priors on the source and lens unlensed trajectory. The first two issues arise due to the implicit assumption that the GEDR3 astrometric solution for both the source and lens does not already contain some of the astrometric microlensing signal. This is a possibility as astrometric microlensing events typically have long tails \citep{Dominik2000, Belokurov2002}, and could overlap with the data used to build the lens and source GEDR3 astrometric solutions. If the lens or source astrometric solution does contain a detectable part of the astrometric microlensing signal, this could potentially bias our inference as the lens and source astrometric parameters would not be representative of the true unlensed lens and source trajectories. 

Firstly, for the source, we have to check that when the GEDR3 data were taken, there was not a significant lensing signal present. As LAWD 37 is so bright \Gaia{} downloads cut-outs and carries out gating so when the background object is within $500$ mas it is downloaded as part of the cut out but with less exposure time, otherwise, standard one dimensional processing is used \citep{Fabricius2016}. GEDR3 is based on data collected between July $2014$ and May $2017$. In May $2017$, the predicted deflection of the source is $<0.2$ mas and the lens source separation is $>>500 $mas. This is below the along scan (AL) precision for a $G\approx18$ mag source \citep[$\sigma_{\text{AL}}\approx0.8$ mas with the standard \Gaia{} pipeline;][]{Rybicki2018,Bramich2018, Everall2021}. We therefore conclude that the astrometric lensing signal was not detectable during the time GEDR3 data were collected and therefore did not significantly influence the GEDR3 astrometric solution of the source. Secondly, for the lens, we have to check if the shift due to the blending with the minor image was detectable during GEDR3 \citep[see Eq. (15) in][]{Bramich2018}. In May $2017$ this shift was $\approx10^{-13}$ mas, we therefore safely conclude that the lens GEDR3 astrometric solution does not contain a significant astrometric lensing signal. 

Finally, we have to consider if the GEDR3 astrometric solution of the $G\approx18$ mag source has likely been influenced by the presence of the comparatively bright $G\approx11$ mag lens. The current \Gaia{} processing pipeline is able to resolve sources for separations in the most optimal cases down to $\approx200$ mas \citep{Arenou2018}. While the lens-source contrast ratio is far from optimal in our case, in May $2017$, the lens and source had a predicted separation of $\approx6400$ mas. At this separation, the lens and source were unlikely to be close to each other on the \Gaia{} focal plane during the GEDR3 time baseline. Therefore, it is safe to conclude that the source GEDR3 astrometric solution is unlikely to have been significantly affected by the presence of the lens. Overall we conclude, for both the lens and source, the GEDR3 solutions are safe to use as priors on the unlensed source and lens trajectories in the models. Finally, it is worth mentioning that the astrometric solution priors from GEDR3 will affect the precision at which we can measure the mass of the lens, we defer discussion of this point to Section \ref{sec:prior_sensitivity}.

\begin{table}
    \begin{tabular}{llp{4cm}}
         Parameter & Prior & Description  \\
         \hline
         $\boldsymbol{\theta}^{\text{ast}}_{S}$ & 
         $\mathcal{N}(\boldsymbol{m}^{\text{G}}_{S}, \boldsymbol{\Sigma}^{G}_{S})$ & GEDR3 prior for the source trajectory \\
          $\boldsymbol{\theta}^{\text{ast}}_{L}$ & $\mathcal{N}(\boldsymbol{m}^{\text{G}}_{L}, \boldsymbol{\Sigma}^{G}_{L})$ & GEDR3 prior for the lens trajectory \\
          $\sigma_{\text{white}}$ & $\mathcal{N}(m_{\sigma_{\text{white}}}, \sigma^{2}_{n})$ & White component of the noise $m_{\sigma_{\text{white}}}=0.8, \sigma_{n}=0.1$ \\
          $\boldsymbol{\sigma}_{\text{corr}}$ & $\mathcal{N}(\boldsymbol{m}_{\text{corr}}, \sigma^{2}_{n}\boldsymbol{I})$ & Correlated components of the noise for each epoch \\
          $\Theta_{\mathrm{E}}$  & $\mathcal{U}(20,60)$ & Flat prior of the angular Einstein radius \\
          \hline
    \end{tabular}
    \caption[LAWD~37 parameter priors]{Summary of the parameter priors used in the models.}
    \label{tab:priors}
\end{table}

\begin{figure}
    \centering
    \includegraphics[width=0.9\columnwidth]{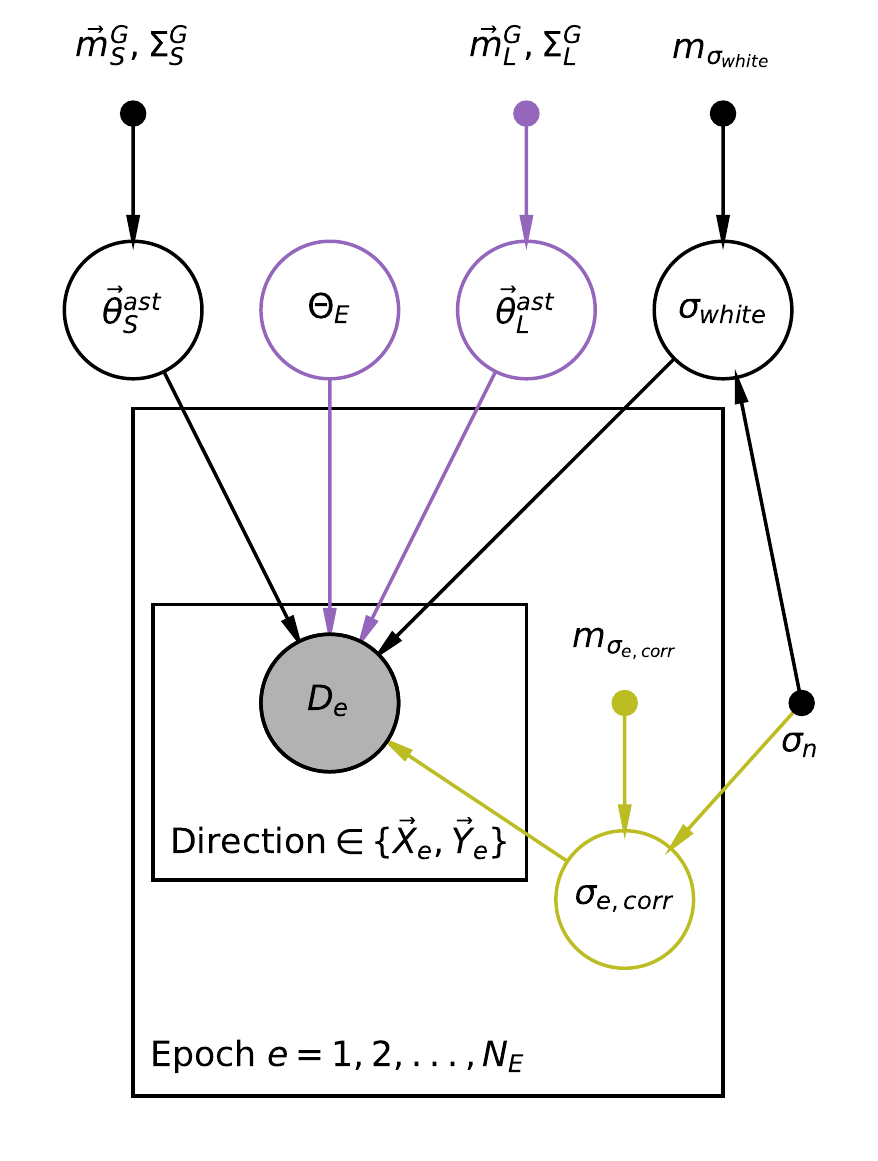}
    \caption[LAWD~37 probabilistic graphical model]{Probabilistic graphical model showing the dependence structure of the models considered in this work. An arrow from one node to another indicates a conditional dependence. No arrow between two nodes means they are conditionally independent. Unfilled circles are latent random variables in the models or parameters that are fitted for. Filled small circles are fixed values in the model (parameters for the informative prior distributions). The shaded circles are the observed data. Parameters inside a plate are repeated for each epoch and then direction. Parameters outside the plates are global parameters. Black parts of the graph are common to all models considered. Purple parts are common to models with an astrometric deflection. Pale green parts of the graph are common to models with correlated noise.}
    \label{fig:dag}
\end{figure}

For the white noise parameter present in all models, we set \\ $p(\sigma_{\text{white}})=\mathcal{N}(\sigma_{\text{white}}|m_{\sigma_{\text{white}}}, \sigma^{2}_{n})$ where $m_{\sigma_{\text{white}}}=0.8$ mas and $\sigma_{n}=0.1$ mas, reflecting the estimated instrument precision of WFC3/UVIS \citep{Bellini2014}. For the correlated noise parameters $\boldsymbol{\sigma}_{\text{corr}}=[\sigma_{1,\text{corr}},\sigma_{2,\text{corr}}, ..., \sigma_{N_{e},\text{corr}}]$ we assume a Gaussian prior $p(\boldsymbol{\sigma}_{\text{corr}})=\mathcal{N}(\boldsymbol{\sigma}_{\text{corr}}|\boldsymbol{m}_{\text{corr}},\sigma^{2}_{n}\boldsymbol{I})$, truncated at zero to avoid negative values. Here, $\boldsymbol{m}_{\text{corr}}=[m_{\sigma_{1,\text{corr}}},m_{\sigma_{2,\text{corr}}}, ...,m_{\sigma_{N_{E},\text{corr}}}]$ are the estimated size of the correlated noise components in Table \ref{tab:corr_noise} and we have assumed no correlation between epochs. Finally, we then assume a uniform prior on $\Theta_{\mathrm{E}}$, as $p(\Theta_{\mathrm{E}})=\mathcal{U}(\Theta_{\mathrm{E}}|\text{lower}=20 \text{ mas},\text{upper}=60 \text{ mas})$. This prior was chosen to be wide enough to be uninformative, yet narrow enough to constrain the model to reasonable areas of the parameter space which allowed fast model fitting. In all models, we build the full prior by taking the product over all the priors of the required parameters. Table \ref{tab:priors} contains a summary of all parameter priors, and Fig. \ref{fig:dag} shows the probabilistic graph illustrating all parameter dependencies and structure of the models. 

\subsection{Sampling the posterior}

Now that we have constructed the prior and likelihoods for each model, we may compute the posterior distribution on the model parameters. The posterior distribution or the probability distribution of the model parameters given the data, is given by Bayes rule,
\begin{equation}
    p(\boldsymbol{\theta} | \mathcal{D}, \mathcal{M}) = \frac{p(\mathcal{D}|\boldsymbol{\theta},\mathcal{M})p(\boldsymbol{\theta}|\mathcal{M})}{p(\mathcal{D}|\mathcal{M})}\propto p(\mathcal{D}|\boldsymbol{\theta},\mathcal{M})p(\boldsymbol{\theta}|\mathcal{M}).
    \label{eq:bayes_rule}
\end{equation}
We obtain samples from the posterior distribution using an MCMC algorithm. Specifically, we use the No-U-Turn Sampler (NUTS) Hamiltonian Monte Carlo algorithm \citep{Homan2014} implemented by the PyMC3 Python package \citep{Salvatier2016}. NUTS allows samples from the posterior distribution to be obtained faster than other traditional MCMC samplers \citep[e.g.,][]{Foreman-Mackey2013} because it uses first-order gradient information to efficiently step through the parameter space. We also take advantage of using the dense full mass matrix step implemented in the Exoplanet Python package for further performance gains \citep{Foreman-Mackey2021}.

For each model investigated in this work, we run NUTS for $2000$ tuning steps and then for a further $10,000$ steps. This is done for two independent chains to permit between-chain convergence checks. Specifically, we compute the rank-normalized $\hat{R}$ convergence diagnostic for each inference in this analysis. $\hat{R}$ measures convergence by comparing between-chain and within-chain variance for each parameter. A value $\hat{R}>1.01$ indicates poor convergence \citep{vehtari2021rank}. We find for all parameters in all inferences considered, $\hat{R}=1.0$, meaning good convergence. Running both chains for a model typically took $10$ Central Processing Unit (CPU) minutes. 

\section{Model comparison and criticism}
\label{sec:comparison}

\subsection{Leave-one-out cross-validation}\label{sec:loo_description}

We are fitting four models to the data. $\mathcal{M}_{DC}$ - a model with the astrometric microlensing deflection and correlated noise, $\mathcal{M}_{DW}$ - a model with the astrometric microlensing deflection and just white noise, $\mathcal{M}_{NC}$ - a model with no astrometric microlensing deflection but with correlated noise, and finally, $\mathcal{M}_{NW}$ - a model with no astrometric microlensing deflection and just white noise. We then need to assess which model best explains the data and critically examine the strengths and weakness of each of the models. To do this, we use the Bayesian Leave-One-Out cross-validation score (LOO). LOO is one method to estimate point-wise out-of-sample prediction accuracy of a given model \citep{Vehtari2017}. LOO is calculated for a given model by fitting the model to the data set where one of the data points has been left out. The posterior samples of that fit are then projected through the model likelihood to assess how well the left-out data point is predicted by the model. The procedure is then repeated so each data point is left out in turn. For a given model, this provides a per data point score which can be totalled over the data to give an indication of overall model performance, or compared data point-wise with a different model allowing an interpretable comparison between models.

Specifically, for a model $\mathcal{M}$, which being fit to the full data set $\mathcal{D}$, the LOO score for the $i$th data point, $D_{i}=\{t_{i},X_{i},Y_{i}\}$, is
\begin{equation}
\text{LOO}_{i, \mathcal{M}} = \log p(D_{i}|\mathcal{D}_{-i}, \mathcal{M}).
\end{equation}
Here $\mathcal{D}_{-i}$ is the full data set $\mathcal{D}$ with the $i$th data point or $D_{i}$ removed. This is the log of the LOO predictive density conditioned on the data set without the $i$th data point. It can be written in terms of expected value of the likelihood of the left-out data point over the posterior distribution obtained while fitting the model to $\mathcal{D}_{-i}$,
\begin{equation}
    p(D_{i}|\mathcal{D}_{-i}, \mathcal{M}) = \int p(D_{i}|\boldsymbol{\theta},\mathcal{D}_{-i},\mathcal{M})p(\boldsymbol{\theta}|\mathcal{D}_{-i},\mathcal{M})d\boldsymbol{\theta}.
    \label{eq:elpd}
\end{equation}
Practically, if we have S samples from the posterior distribution of the model parameters $\{\boldsymbol{\theta}^{s}_{-i}\}_{s=1}^{S}$, obtained by fitting model $\mathcal{M}$ to the data set $\mathcal{D}_{-i}$, we can use these samples to compute Eq. (\ref{eq:elpd}) as,
\begin{equation}
    p(D_{i}|\mathcal{D}_{-i},\mathcal{M})= \frac{1}{S}\sum_{s=1}^{S}p(D_{i}|\mathcal{D}_{-i},\boldsymbol{\theta}^{s}_{-i}, \mathcal{M}).
\end{equation}
Here, we have assumed that we have a sufficient number of samples to fully capture the posterior distribution \citep{Gelman2014}. 

This quantity can be totaled over the data to give an overall score for model $\mathcal{M}$, 
\begin{equation}
    \text{LOO}_{\mathcal{M}} = \sum_{i=1}^{N}\text{LOO}_{i, \mathcal{M}}.
    \label{eq:total_loo}
\end{equation}
Alternatively the difference can be used to compare the performance of two models $\mathcal{M}_{1}$ and $\mathcal{M}_{2}$,
\begin{equation}
    \Delta\text{LOO}_{\mathcal{M}_{1},\mathcal{M}_{2}} = \text{LOO}_{\mathcal{M}_{1}} - \text{LOO}_{\mathcal{M}_{2}}.
    \label{eq:loo_diff}
\end{equation}
Here, a positive value indicates that $\mathcal{M}_{1}$ has a higher computed out-of-sample predictive accuracy than $\mathcal{M}_{2}$. The standard error (se), on this difference is given by \cite{Vehtari2017} as,
\begin{equation}
    \text{se}\left(\text{LOO}_{\mathcal{M}_{1},\mathcal{M}_{2}}\right) = \sqrt{N\text{V}_{i=1}^{N}(\text{LOO}_{i, \mathcal{M}_{1}}-\text{LOO}_{i, \mathcal{M}_{2}})}.
    \label{eq:se_diff}
\end{equation}
$V_{i=1}^{N}$ is the variance of the point-wise difference over the full data set of N data points. This can be used to assess how significant the difference between two models is. For brevity, we define the significance of the difference as  $\text{sig}(\bullet)= |\bullet| / \text{se}(\bullet)$ where $\bullet = \text{LOO}_{\mathcal{M}_{1},\mathcal{M}_{2}}$.

In the case of this analysis, it is also informative to sum over the point-wise LOO score over the data in a single epoch, $e$. This is because the data are tightly temporally clustered within an epoch, and in the correlated noise models ($\mathcal{M}_{DC},\mathcal{M}_{NC}$), data within an epoch share correlated noise properties. We define the difference in LOO predictive accuracy over an epoch $e$, as $\Delta\text{LOO}^{e}_{\mathcal{M}_{1},\mathcal{M}_{2}}$. This quantity is analogous to  $\Delta\text{LOO}_{\mathcal{M}_{1},\mathcal{M}_{2}}$ (defined in Eqs. \ref{eq:total_loo} and \ref{eq:loo_diff}), but instead of summing the point-wise score over all data points, the sum is taken only over the data in epoch, $e$. The standard error, $\text{se}(\Delta\text{LOO}^{e}_{\mathcal{M}_{1},\mathcal{M}_{2}})$, is also calculated analogously by instead calculating the variance of the point-wise difference of data within the epoch. In this case, $N$ in Eq. \ref{eq:se_diff} is the total number of data points within epoch $e$. Overall, computation of $\Delta\text{LOO}^{e}_{\mathcal{M}_{1},\mathcal{M}_{2}}$ will permit the comparison of models at epoch resolution.

Calculating all $\text{LOO}_{i,\mathcal{M}}$ terms for all models is computationally expensive. This is because it would require $N$ full refits (obtaining samples from the posterior distribution) of the model with each data point left out in turn. In our case, $N=81$ and a single refit of a model takes $\approx 10$ CPU minutes, therefore computing all $\text{LOO}_{i,\mathcal{M}}$ terms for the four considered models would take $\approx 55$ CPU hours of computation. While not completely unfeasible, the required computation is still significant. We therefore turn to an importance sampling approximation to compute the $\text{LOO}_{i,\mathcal{M}}$ terms for each model.

We use the Pareto Smoothed Importance Sampling \citep[PSIS;][]{Vehtari2015} approximation to compute the $\text{LOO}_{i,\mathcal{M}}$ terms for each model \citep{Vehtari2017, Burkner2020}, implemented in the Arviz Python package \citep{arviz}. Instead of refitting the model with each data point left out in turn, in the PSIS approximation the model is initially fit to the full data set. Then posterior samples from the full data set fit are re-weighted (via importance sampling) to approximate the effect of removing each data point in turn. Overall PSIS allows the fast and approximate computation of the LOO terms for a model with very few refits. Appendix \ref{sec:loo} contains the application details of the PSIS approximation along with checks of the approximation accuracy for the models considered in this work.

\subsection{The case for LOO over other comparison metrics}

LOO is just one of many metrics that can be computed to assess the performance of a model. In this section, we briefly justify the decision to use LOO as a model criticism and comparison tool compared to the two commonly used approaches in astronomy: a reduced $\Delta\chi^{2}$ approach, and use of the Bayesian model evidence.

Typically in microlensing event analyses, albeit in analyses of photometric microlensing events, a reduced $\Delta\chi^{2}$ approach is used to select between competing models \citep[e.g.,][]{Bond2004, Smith2005,Alcock2000, Bennett2018}. There is a multitude of reasons why we do not use it here and choose LOO instead. Firstly, because some of the models considered in this work contain Gaussian-correlated noise components, reduced $\Delta\chi^{2}$ is no longer fully descriptive of the likelihood of the model (reduced $\Delta\chi^{2}$ is only related to the likelihood of an uncorrelated Gaussian likelihood with a diagonal covariance matrix). Secondly, reduced $\Delta\chi^{2}$ is only valid for a model that is linear in its parameters; none of the models considered in this work are linear in all their parameters. Thirdly, reduced $\Delta\chi^{2}$ fails to account for any posterior uncertainty on the parameters. \cite{Andrea2010} gives an extensive account of the pitfalls of using reduced $\Delta\chi^{2}$ for comparison of non-linear models.
\begingroup
\renewcommand{\arraystretch}{1.3}
\setlength{\tabcolsep}{3pt}
\begin{table*}
    \begin{tabular}{l|lllll}
    Parameter & Unit &  \multicolumn{4}{c}{Model} \\
     & & $\mathcal{M}_{DC}$ & $\mathcal{M}_{DW}$ & $\mathcal{M}_{NC}$ & $\mathcal{M}_{NW}$ \\
    \hline
    $X_{0,S}$ & mas & $35186.997^{0.127}_{-0.128}$ & $35187.027^{0.124}_{-0.123}$ & $35186.943^{0.126}_{-0.125}$ & $35186.946^{0.122}_{-0.125}$ \\
$Y_{0,S}$ & mas &$45709.705^{0.125}_{-0.124}$ & $45709.675^{0.121}_{-0.122}$ & $45709.726^{0.125}_{-0.124}$ & $45709.741^{0.123}_{-0.122}$ \\
$\mu_{X,S}$ & mas/year& $8.964^{0.044}_{-0.043}$ & $8.992^{0.040}_{-0.040}$ & $9.011^{0.045}_{-0.045}$ & $9.042^{0.045}_{-0.044}$ \\
$\mu_{Y,S}$ & mas/year & $0.028^{0.047}_{-0.047}$ & $-0.012^{0.046}_{-0.047}$ & $0.159^{0.046}_{-0.046}$ & $0.289^{0.043}_{-0.044}$ \\
$\pi_{S}$ & mas & $0.123^{0.110}_{-0.111}$ & $0.043^{0.101}_{-0.098}$ & $0.127^{0.121}_{-0.121}$ & $0.070^{0.115}_{-0.114}$ \\
$\sigma_{\text{white}}$ & mas & $0.838^{0.044}_{-0.041}$ & $0.993^{0.046}_{-0.044}$ & $0.882^{0.046}_{-0.045}$ & $1.225^{0.051}_{-0.048}$ \\
$X_{0,L}$ & mas & $45825.738^{0.014}_{-0.014}$ & $45825.738^{0.013}_{-0.014}$ &  - & - \\
$Y_{0,L}$ & mas & $46592.483^{0.015}_{-0.015}$ & $46592.483^{0.015}_{-0.015}$ &  - & - \\
$\mu_{X,L}$ & mas/year & $-2661.640^{0.018}_{-0.018}$ & $-2661.640^{0.018}_{-0.018}$ &  - & - \\
$\mu_{Y,L}$ & mas/year & $-344.932^{0.019}_{-0.019}$ & $-344.932^{0.020}_{-0.019}$ &  - & - \\
$\pi_{L}$ & mas &$215.675^{0.018}_{-0.018}$ & $215.676^{0.018}_{-0.018}$ &  - & - \\
$\Theta_{\mathrm{E}}$ & mas & $31.353^{2.077}_{-2.184}$ & $34.164^{1.386}_{-1.440}$ &  - & - \\
$\sigma_{{1,\text{corr}}}$ & mas & $0.081^{0.077}_{-0.055}$& - & $0.084^{0.083}_{-0.057}$& - \\
$\sigma_{{2,\text{corr}}}$ & mas & $0.127^{0.090}_{-0.075}$& - & $0.206^{0.097}_{-0.103}$& - \\
$\sigma_{{3,\text{corr}}}$ & mas  &  $0.597^{0.097}_{-0.099}$& - & $0.627^{0.096}_{-0.094}$& - \\
$\sigma_{{5,\text{corr}}}$ & mas & $0.620^{0.096}_{-0.093}$& - & $0.741^{0.082}_{-0.080}$& - \\
$\sigma_{{6,\text{corr}}}$ & mas & $0.380^{0.098}_{-0.099}$& - & $0.477^{0.087}_{-0.083}$ & - \\
$\sigma_{{7,\text{corr}}}$ & mas & $0.681^{0.087}_{-0.085}$& - & $0.719^{0.083}_{-0.081}$& - \\
$\sigma_{{8,\text{corr}}}$ & mas & $0.078^{0.077}_{-0.053}$& - & $0.120^{0.094}_{-0.078}$& - \\
$\sigma_{{9,\text{corr}}}$ & mas & $0.085^{0.078}_{-0.057}$& - & $0.105^{0.090}_{-0.069}$& - \\
\hline
    \end{tabular}
    \caption[LAWD 37 posterior summaries]{Parameter posterior summaries for each model. Values are the posterior medians, uncertainties are the $84$th-$50$th and $50$th-$16$th posterior percentiles. '-' indicates that the considered model does not contain the parameter.}
    \label{tab:posterior}
\end{table*}
\endgroup

Comparison of models using the Bayesian evidence (the denominator in Eq.~\ref{eq:bayes_rule}) is becoming popular in astrophysics due to nested sampling algorithms that readily allow its computation \citep[e.g.,][]{Skilling2006, Higson2019, Speagle2020}. The Bayesian evidence has the appealing properties of fully capturing parameter uncertainty and naturally penalizes more complex models that do not significantly explain the data better. The critical downside, however, is that the evidence is sensitive to the choice of prior distribution \citep[see e.g.,][]{Fong2020}. This becomes a problem when comparing models possessing parameters with uninformative and somewhat arbitrarily set prior distributions \citep[see e.g., Section 7.2 of][]{Gelman2013book}. For example, for $\Theta_{\mathrm{E}}$ in this analysis, the arbitrary choice of a large width for its uninformative prior can arbitrarily change the model evidence without any resulting change of the posterior distribution. 

Comparatively, LOO is typically less sensitive to the model priors. This is because each $\text{LOO}_{i,\mathcal{M}}$ term is computed with the model conditioned on the rest of data (Eq. \ref{eq:elpd}). This means in scenarios where the number of data points is large, the prior can be overwhelmed by the likelihood and has less of an effect on each $\text{LOO}_{i,\mathcal{M}}$ term. In the analysis presented in this work, although the number of fitted data points is $81$, it is difficult to asses if this constitutes a large data set due to the use of informative priors on the source and lens astrometry, and the noise components of the models. Finally, the Bayesian evidence only provides a single summary statistic for the whole model and data, shedding little light on precisely where a model fails, whereas LOO provides an interpretable per data point score.

\section{Results}
\label{sec:results_lawd37}

\subsection{Performance of the models}\label{sec:performace}

\begin{figure*}
    \centering
    \includegraphics[width=1.0\textwidth]{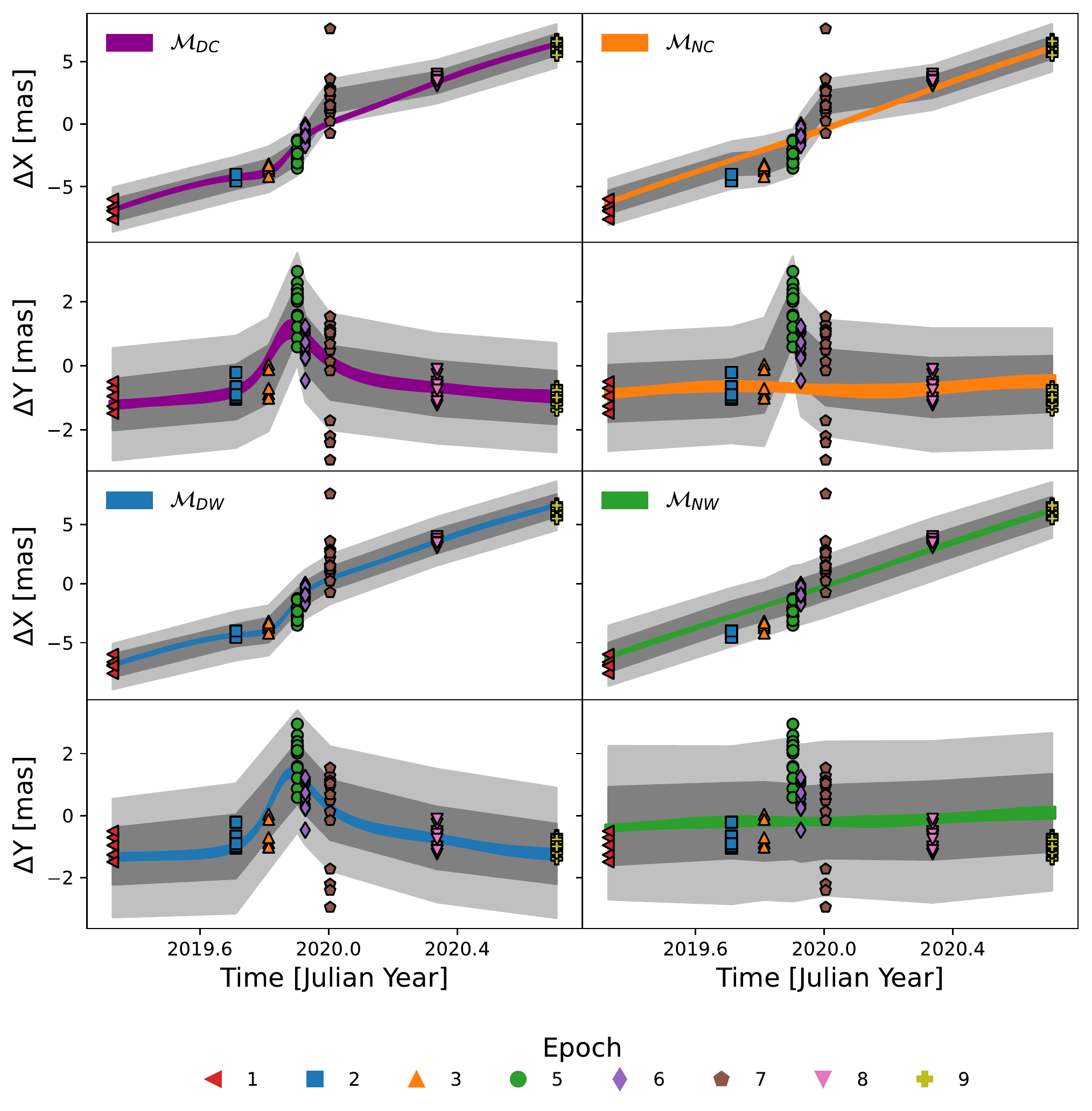}
    \caption[LAWD 37 posterior realizations over data]{Posterior realizations plotted over the data in the X and Y directions and for each of the considered models. Coloured bands show the $84$th-$16$th posterior percentiles on the inferred source trajectory for each model and direction. Dark (light) grey bands are the posterior data realizations $16$th-$84$th ($2$nd-$98$th) percentile bands. Specifically this includes the trajectory and the noise model component realizations. The posterior data realizations are discontinuous between epoch because the noise model is only defined within an epoch and on the data grid.}
    \label{fig:posterior_over_data}
\end{figure*}

All models ($\mathcal{M}_{DC}$, $\mathcal{M}_{DW}$, $\mathcal{M}_{NC}$, $\mathcal{M}_{NW}$) were fitted to the data. Posterior parameter summaries for all parameters can be found in Table \ref{tab:posterior}, and posterior projections of the model over the data are shown in Fig. \ref{fig:posterior_over_data}. Additionally, LOO scores were computed for all models as described in Section \ref{sec:loo_description}.

Table \ref{tab:loo_matrix} shows the pairwise comparison between all model combinations and the total LOO scores. $\mathcal{M}_{DC}$ (deflection and correlated noise model) has the best overall LOO score when compared to all other models. $\mathcal{M}_{NW}$ is the least preferred model with all other models having a higher score. The model most competitive with $\mathcal{M}_{DC}$, is $\mathcal{M}_{NC}$ with $\Delta\text{LOO}_{\mathcal{M}_{DC},\mathcal{M}_{NC}}=8.6$ and $\text{sig}(\Delta\text{LOO}_{\mathcal{M}_{DC},\mathcal{M}_{NC}})=1.5$. This means that the correlated component of the noise is an important feature of the model, since the deflection model with just white noise ($\mathcal{M}_{DW}$) is comparably not competitive with $\mathcal{M}_{NC}$. In fact, Table \ref{tab:loo_matrix} shows that $\mathcal{M}_{NC}$ is preferred over $\mathcal{M}_{DW}$ with $\Delta\text{LOO}_{\mathcal{M}_{NC},\mathcal{M}_{DW}}=-20.9$ and $\text{sig}(\Delta\text{LOO}_{\mathcal{M}_{NC},\mathcal{M}_{DW}})=2.2$. That is, the non-deflection correlated noise model is preferred over the model with the deflection and just white noise. While $\mathcal{M}_{DC}$ is the overall preferred model, it is informative to understand how exactly $\mathcal{M}_{NC}$ is able to explain the data with no deflection term and even beat the deflection model with just white noise, $\mathcal{M}_{DW}$.

\begin{figure}
    \centering
    \includegraphics[width=\columnwidth]{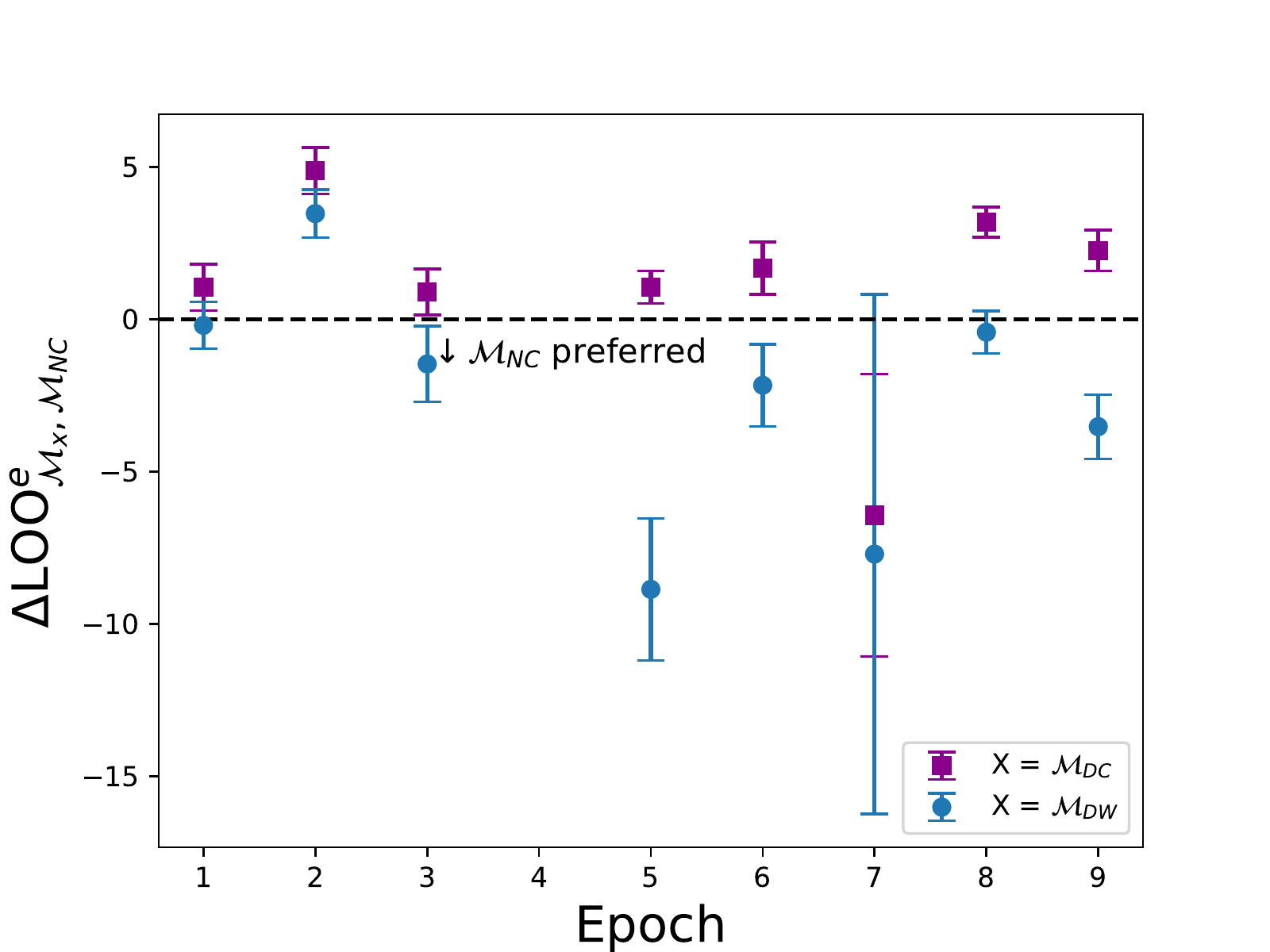}
    \caption[Epoch model comparison]{LOO score of models $\mathcal{M}_{DC}$ and $\mathcal{M}_{DW}$ compared to $\mathcal{M}_{NC}$ within each epoch. Error bars are one standard error.}
    \label{fig:epoch_mean_loo}
\end{figure}

The starting point for understanding the good performance of $\mathcal{M}_{NC}$ is to examine the per-epoch LOO scores. Fig. \ref{fig:epoch_mean_loo} shows the per-epoch LOO scores for $\mathcal{M}_{DW}$ and $\mathcal{M}_{DC}$ compared to $\mathcal{M}_{NC}$. For the comparison of the two correlated noise models ($\mathcal{M}_{DC}$ versus $\mathcal{M}_{NC}$), it is shown that $\mathcal{M}_{DC}$ marginally beats $\mathcal{M}_{NC}$ in every epoch with the exception of epoch $2$ where $\mathcal{M}_{DC}$ is more clearly preferred and epoch $7$ where $\mathcal{M}_{NC}$ beats $\mathcal{M}_{DC}$. The reason why $\mathcal{M}_{NC}$ can explain the epoch with the largest deflection terms (epochs $5,6$, and $7$), is that it inflates the correlated noise and alters the unlensed source trajectory. 

\begin{table}
    \centering
    $\Delta \text{LOO}_{\text{row},\text{column}}$
    \begin{tabular}{l|llll}
          & $\mathcal{M}_{DC}$ & $\mathcal{M}_{DW}$ & $\mathcal{M}_{NC}$ & $\mathcal{M}_{NW}$ \\
         \hline
         $\mathcal{M}_{DC}$ & - & $29.5(3.5)$ & $8.6(1.5)$ & $76.5(4.6)$ \\
         $\mathcal{M}_{DW}$ & $-29.5(3.5)$ & - & $-20.9(2.2)$ & $47.1(3.4)$\\
         $\mathcal{M}_{NC}$ & $-8.6(1.5)$ & $20.9(2.2)$ & - &  $67.9(5.4)$\\
         $\mathcal{M}_{NW}$ & $-76.5(4.6)$ & $-47.1(3.4)$  & $-67.9(5.4)$& -\\
         \hline
         LOO$_{\mathcal{M}}$ & $-215.6$ & $-245.0$ & $-224.1$ & $-292.1$\\
         se(LOO$_{\mathcal{M}}$) & $38.3$ & $37.8$ & $33.2$ & $26.0$ \\
         \hline
    \end{tabular}
    \caption{Pairwise difference in LOO scores for all models considered. Positive number means the model in the row is preferred. Significance, sig($\Delta\text{LOO}_{\text{row},\text{column}}$), of the scores are indicated in with the parentheses. Total LOO score for each more and its standard error are shown in the bottom two rows. In order of descending LOO score (highest to lowest predictive accuracy) the models are $\mathcal{M}_{DC}$, $\mathcal{M}_{NC}$, $\mathcal{M}_{DW}$ and $\mathcal{M}_{NW}$.}
    \label{tab:loo_matrix}
\end{table}

%\begin{figure}
%    \centering
%    \includegraphics[width=\columnwidth]{figs/loo_matrix.pdf}
%    \caption[LAWD 37 model comparison]{\textbf{Left}: Pairwise difference in LOO scores for all models considered. Red or a positive number means the model in the column is preferred. Blue or a negative number means the model in the row is preferred. \textbf{Right}: Significance of the scores in the left panel. Specifically, each score in the left panel divided by its standard error estimate. A higher number means the difference is more significant.}
%    \label{fig:loo_matrix}
%\end{figure}

\begin{figure*}
    \centering
    \includegraphics[width=1.0\textwidth]{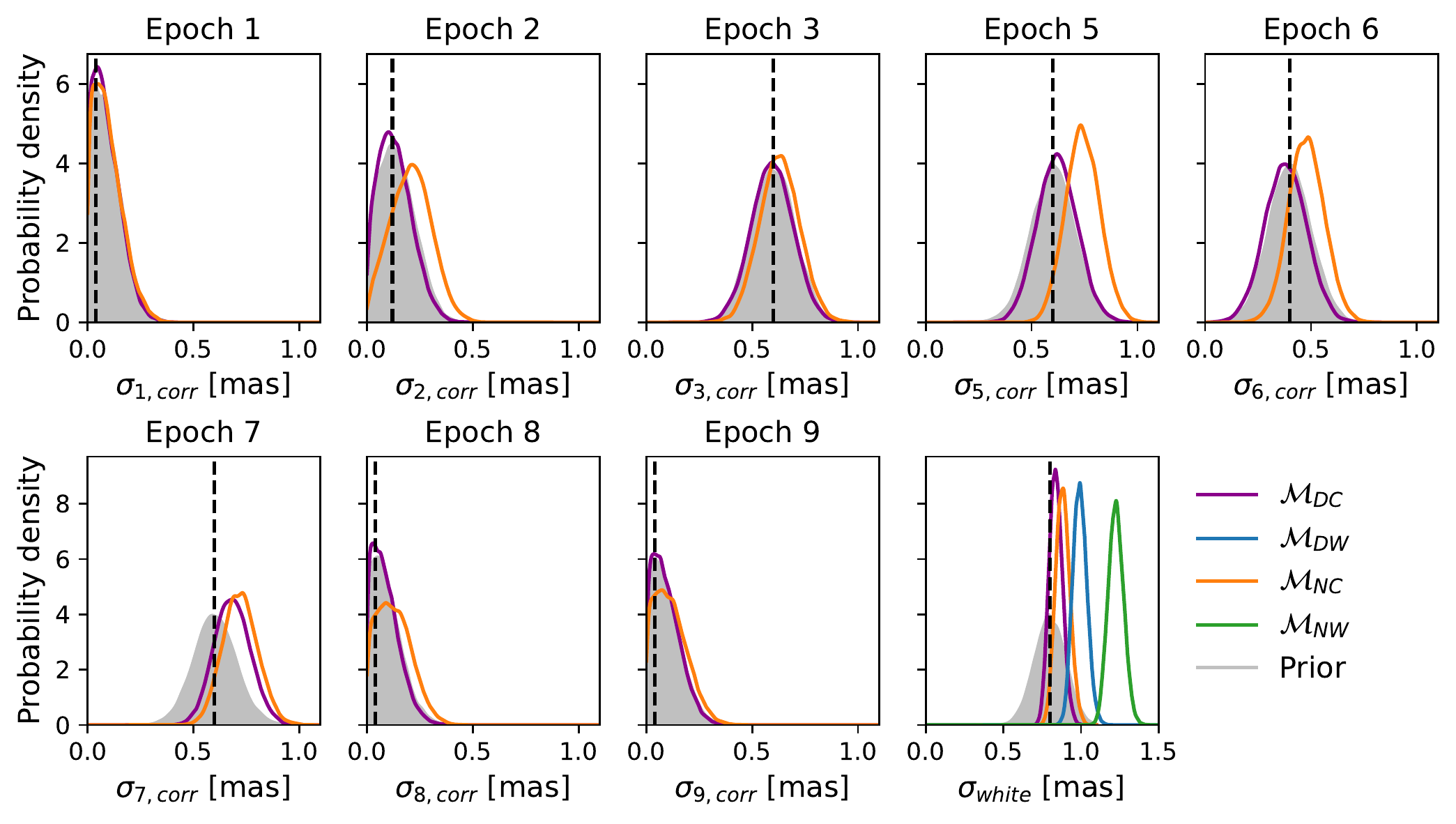}
    \caption[Prior versus posterior on noise parameters]{Prior and marginal posterior probability density functions for the noise parameters in the different models. Vertical dashed lines show the values of these parameters estimated from the PSF subtraction simulations (Table \ref{tab:corr_noise}). The prior probability density function is shaded to aid differentiation with the posteriors. Note that the models with no correlated noise, $\mathcal{M}_{DW}$ and $\mathcal{M}_{NW}$, do not have correlated noise parameters and therefore, only the posteriors on $\sigma_{\text{white}}$ are shown.}
    \label{fig:noise_prior_vs_posterior}
\end{figure*}

Fig. \ref{fig:noise_prior_vs_posterior} shows the priors and posterior distributions of the noise parameter in all of the models. For the correlated noise parameters, $\mathcal{M}_{NC}$ inflates the size of the noise parameters relative to $\mathcal{M}_{DC}$ and the prior for epochs $5,6$ and $7$. $\mathcal{M}_{NC}$ does this to try and explain the deflection signal. For epoch $2$, the correlated noise term in $\mathcal{M}_{NC}$ is slightly inflated compared to $\mathcal{M}_{DC}$ and the prior. At first glance, this seems counterintuitive because the deflection at epoch $2$ is small. This begs the question as to what signal $\mathcal{M}_{NC}$ is trying to explain away with increased correlated noise. The reason for this is that, although the signal at epoch $2$ is small, so is the estimated magnitude of the correlated noise, and so epoch $2$ has one of the highest signal-to-noise ratios of all the epochs. Overall, this means correlated noise can mimic the deflection signal. It is noted that in epoch $7$ and for the white noise both $\mathcal{M}_{NC}$ and $\mathcal{M}_{DC}$ inflate the noise terms relative to the prior. This suggests that the priors underestimate both of these quantities. It is also noted that for models missing either the correlated noise or the deflection signal, the white noise is increased to compensate relative to $\mathcal{M}_{DC}$ and the prior.

\begin{figure}
    \centering
    \includegraphics[width=\columnwidth]{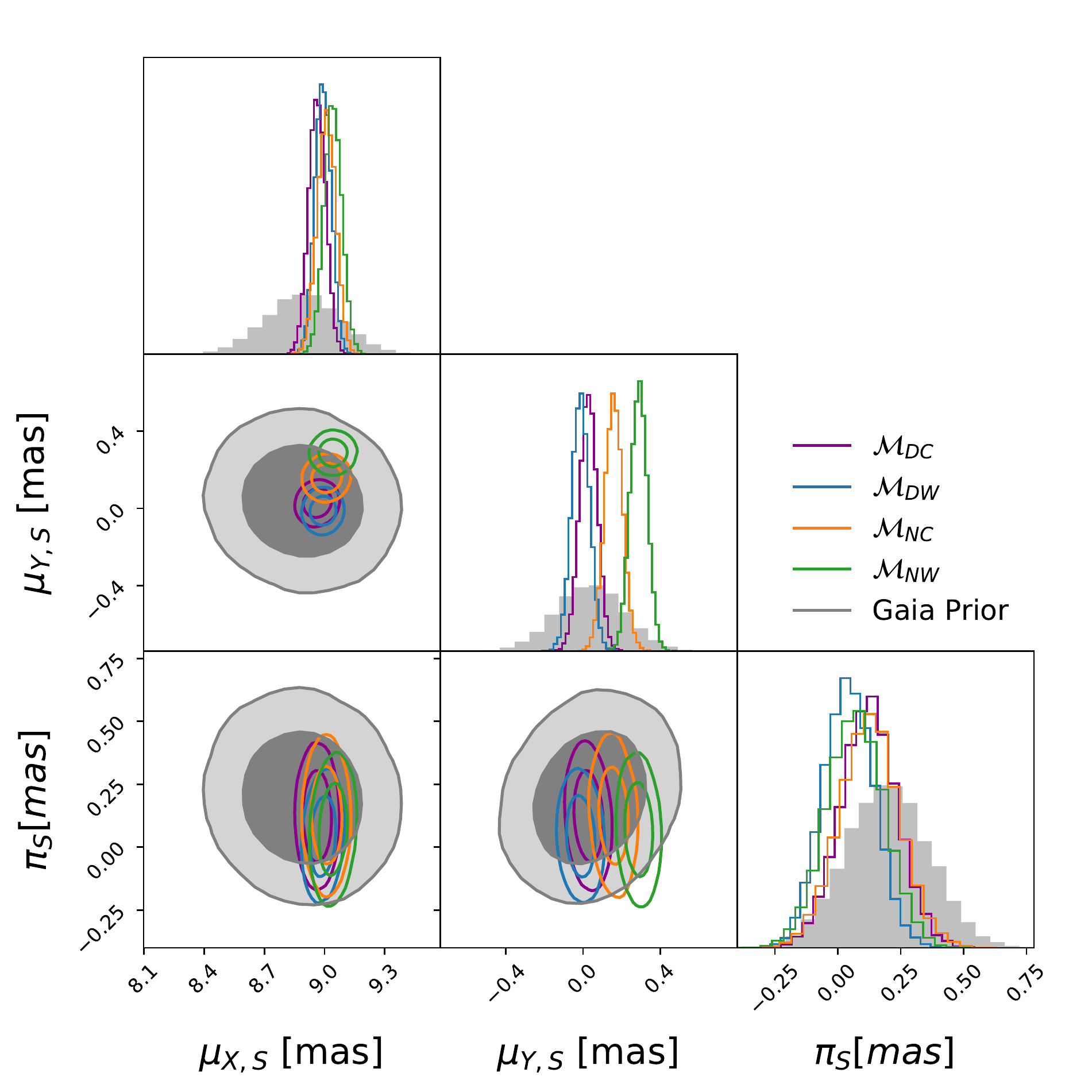}
    \caption[LAWD 37 source astrometry posterior]{Comparison of the source GEDR3 astrometry used as a prior in all of the models compared with the posterior on the astrometry from each of the models. All panels show a probability density. For the 2D plots, the inner and outer contours contain $68\%$ and $95\%$ of the probability mass, respectively. The histograms show the marginal probability densities for each parameter. We have omitted the GEDR3 prior and model posteriors for the source reference positions ($X_{0,S}$, $Y_{0,S}$) because we found good agreement between GEDR3 priors and all the model posteriors (see Fig. \ref{fig:full_source_ast} for the full corner plot).}
    \label{fig:source_prior_vs_posterior}
\end{figure}

Fig. \ref{fig:source_prior_vs_posterior} shows the GEDR3 priors and posterior distributions of the source astrometric parameters. While there is broad agreement between all models for the source proper motion in the $X$ direction and the source parallax, $\mathcal{M}_{NC}$ and $\mathcal{M}_{DC}$ disagree on the Y direction proper motion posterior. $\mathcal{M}_{DC}$ infers $\mu_{Y,S}=0.028^{+0.047}_{-0.047}$ mas/year, whereas $\mathcal{M}_{NC}$ infers $\mu_{Y,S}=0.159^{+0.046}_{-0.046}$ mas/year. The relatively high value inferred by $\mathcal{M}_{NC}$ is caused by $\mathcal{M}_{NC}$ trying to explain away the positive Y direction deflection (see e.g., Fig. \ref{fig:posterior_over_data}) by altering the source trajectory. This means that further data taken after the event to pin down $\mu_{Y,S}$ could completely rule out $\mathcal{M}_{NC}$ as a viable model. Encouragingly, we note that for all models the lens and source astrometric parameters are consistent with the GEDR3 priors (see Figs. \ref{fig:full_lens_ast} and \ref{fig:full_source_ast} in Appendix \ref{sec:consitency_with_gaia}).

The reason for the better performance of $\mathcal{M}_{DW}$ compared with $\mathcal{M}_{NC}$ in epoch $2$ is a high signal-to-noise deflection (as mentioned earlier), combined with the inflexibility of the source trajectory to be altered to explain away an offset in the negative $X$ direction. This is due to the asymmetrical deflection in the $X$ direction being in the negative (before the closest approach) then positive (after the closest approach) $X$ direction. The source trajectory cannot be altered in $\mathcal{M}_{NC}$ to account for both of these offsets, so $\mathcal{M}_{NC}$ performs worse than $\mathcal{M}_{DC}$ in epoch $2$. The better performance of $\mathcal{M}_{NC}$ compared to $\mathcal{M}_{DW}$ in epoch $7$ is due to the outlying data within that epoch (see Fig. \ref{fig:posterior_over_data}). These outlying data are located at lower $Y$ values which are further away from the deflection trajectory of $\mathcal{M}_{DC}$ than the unlensed trajectory of $\mathcal{M}_{NC}$.

For the per-epoch comparison of $\mathcal{M}_{NC}$ and $\mathcal{M}_{DW}$, Fig. \ref{fig:epoch_mean_loo} shows that $\mathcal{M}_{NC}$ beats $\mathcal{M}_{DW}$ in every epoch apart from epoch $2$. The relative performance of $\mathcal{M}_{NC}$ and $\mathcal{M}_{DW}$ can be explained by the same reasoning used above, for the $\mathcal{M}_{DC}$ and $\mathcal{M}_{NC}$ epoch comparison. In epoch $5$, Fig. \ref{fig:epoch_mean_loo} shows $\mathcal{M}_{DW}$ clearly performs worse than $\mathcal{M}_{NC}$, despite there being a large deflection signal in epoch $5$. This is due to epoch $5$ also having a large correlated noise estimate ($m_{\sigma_{e,\text{corr}}}=0.6$ mas) which white noise $\mathcal{M}_{DW}$ cannot explain away, even by inflating the size of the white noise (see Fig. \ref{fig:noise_prior_vs_posterior}).

Fig. \ref{fig:data} shows that epoch 7 has a large scatter in the data compared with all of the other epochs. This is due to the source lying close to a column bleed as shown in Fig. \ref{fig:lawd_74_mosaic}, which meant a specialized measuring procedure to fit only the uncorrupted pixels had to be used. This could mean that the data in epoch 7 are particularly unreliable. Furthermore, because all of the data in epoch 7 are potentially effected by the source's proximity to the charge bleed column a LOO analysis is not sensitive to this, since withholding one epoch 7 data point leaves in the remaining 12. Therefore as a safety check we compare the two best models $\mathcal{M}_{DC}$ and $\mathcal{M}_{NC}$ whilst leaving all of epoch 7 out of the analysis. In this case $\text{LOO}_{\mathcal{M}_{DC}}=-95.2\pm7.9$ and $\text{LOO}_{\mathcal{M}_{NC}}=-103.8\pm7.2$. This means the model with the deflection and correlated noise is still preferred over $\mathcal{M}_{NC}$, but with slightly higher significance, sig($\Delta$LOO)$=2.7$. Overall, the model selection conclusions are not sensitive to including or withholding of epoch 7 for $\mathcal{M}_{DC}$ and $\mathcal{M}_{NC}$.

\subsection{Inference on the angular Einstein radius}\label{sec:theta_E_inferece}

\begin{figure}
    \centering
    \includegraphics[width=\columnwidth]{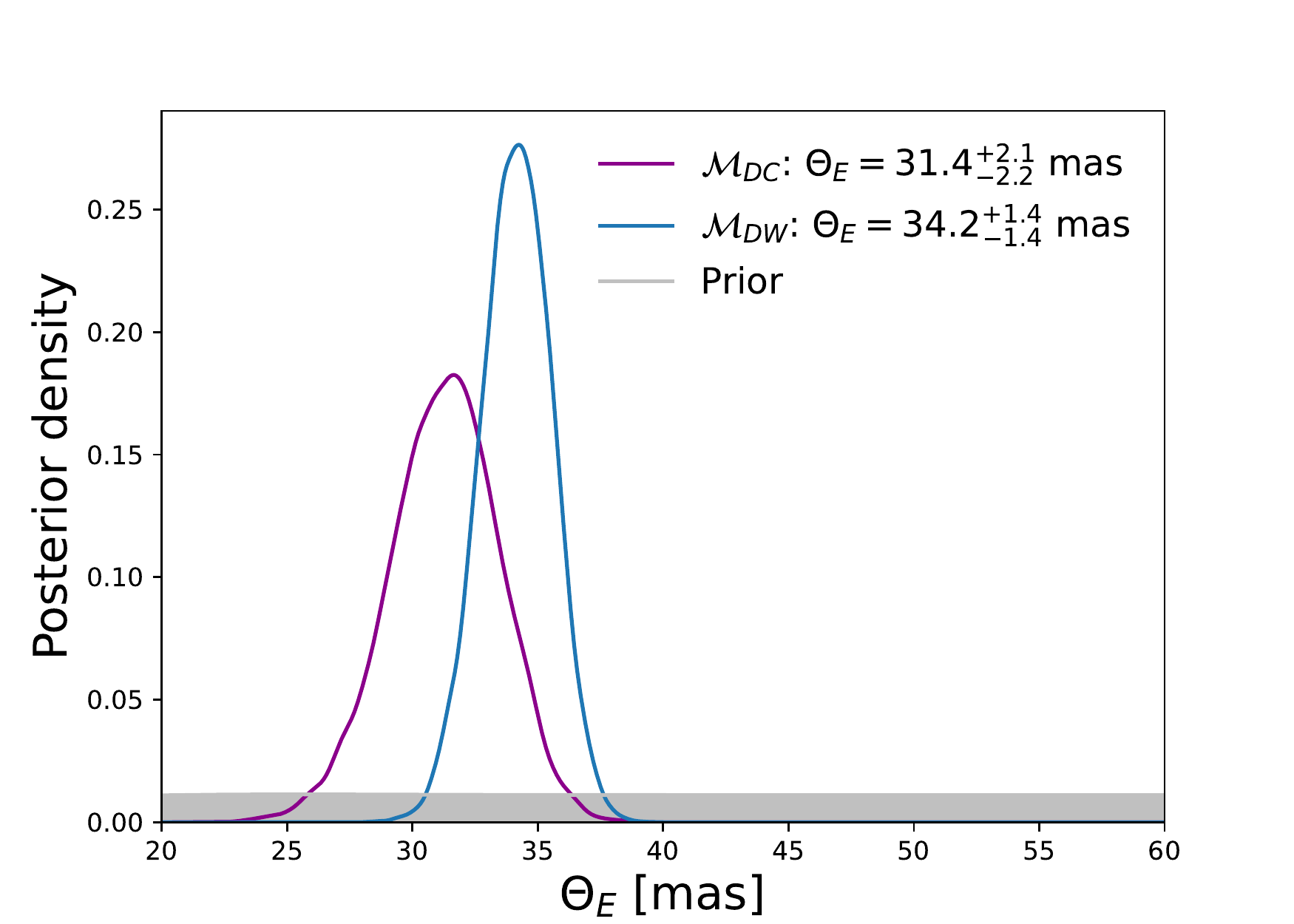}
    \caption[LAWD 37 angular Einstein radius posterior]{Posterior distribution on $\Theta_{\mathrm{E}}$ for the deflection model with correlated noise $\mathcal{M}_{DC}$ and the deflection model with white noise $\mathcal{M}_{DW}$. Reported $\Theta_{\mathrm{E}}$ values are the $50$th posterior percentile and the $84$th-$16$th posterior percentile uncertainties. The prior used for $\Theta_{\mathrm{E}}$ in both models is shown in grey.}
    \label{fig:theta_E}
\end{figure}

Both models including the astrometric lensing deflection signal ($\mathcal{M}_{DW}$ and $\mathcal{M}_{DC}$), provide a posterior inference on $\Theta_{\mathrm{E}}$. Fig. \ref{fig:theta_E} shows the $\Theta_{\mathrm{E}}$ marginal posterior distribution for both the $\mathcal{M}_{DW}$ and $\mathcal{M}_{DC}$ models, along with the prior used in both models. For $\mathcal{M}_{DW}$ and $\mathcal{M}_{DC}$, the inferred values are $\Theta_{\mathrm{E}}=34.2^{+1.4}_{-1.4}$ mas and $\Theta_{\mathrm{E}}=31.4^{+2.1}_{-2.2}$ mas, respectively. Here the values and upper and lower error bars represent the $50$th, $84$th-$50$th, and $16$th-$50$th posterior percentiles, respectively. Fig. \ref{fig:theta_E} shows that $\mathcal{M}_{DW}$ provides a tighter constraint and slightly higher value of $\Theta_{\mathrm{E}}$ compared with $\mathcal{M}_{DC}$. It is also shown that the $\mathcal{M}_{DC}$ $\Theta_{\mathrm{E}}$ posterior distribution is slightly asymmetrical with more probability mass towards lower $\Theta_{\mathrm{E}}$ values.

The difference in the $\Theta_{\mathrm{E}}$ posteriors between $\mathcal{M}_{DW}$ and $\mathcal{M}_{DC}$ is consistent with the findings in Section \ref{sec:comparison}. In Section \ref{sec:comparison}, it was shown that the correlated noise part of the model (with help from a positive $\mu_{Y,S}$) can mimic or explain away some of the microlensing signal. Therefore, jointly fitting both the correlated noise and the microlensing signal means the observed offset on position explanation is shared amongst the microlensing signal and correlated noise, causing a slightly lower posterior median $\Theta_{\mathrm{E}}$ (as $\delta_{+}\propto\Theta_{\mathrm{E}}$ for $u\gg1$) for $\mathcal{M}_{DC}$ compared with $\mathcal{M}_{DW}$. Conversely, because the white noise model in $\mathcal{M}_{DW}$ cannot explain correlated noise at each epoch, it inflates the values of $\Theta_{\mathrm{E}}$ to compensate. The larger spread in the $\Theta_{\mathrm{E}}$ posterior for $\mathcal{M}_{DC}$ is likely similarly due to the fact the correlated noise can mimic the microlensing signal, mean that there is some degeneracy between them, which ultimately leads to a less certain inference on $\Theta_{\mathrm{E}}$ in $\mathcal{M}_{DC}$.

We now use the posterior samples to calculate an inferred value for $M_{\text{L}}$, the mass of LAWD 37. Inverting the expression for $\Theta_{\mathrm{E}}$, we can write the $M_{\text{L}}$ in terms $\Theta_{\mathrm{E}}$ and $\pi_{L}-\pi_{S}$. Of course, we run into the same difficulties described in Section \ref{sec:parametisation} of negative $\pi_{S}$ values entering as distance terms in $\Theta_{\mathrm{E}}$. This is not a problem because the source is so distant, $\pi_{S}\approx0.1$ mas, compared with the lens, $\pi_{L}\approx215$ mas, we can approximate $\pi_{L}-\pi_{S}\approx\pi_{L}$.  Under this approximation we get $M_{\text{L}}$ for the correlated noise model, $M_{L}=0.56\pm0.08 M_{\odot}$, and $M_{\text{L}}$ for the white noise model, $M_{L}=0.66^{+0.06}_{-0.05} M_{\odot}$. Here, we report the $50$th, $84$th-$16$th, and $50$th-$16$th posterior percentiles.  We note inclusion of $\pi_{S}$ both with negative posterior samples and with the negative posterior values truncated, does not change the reported mass values or uncertainties for either model.

We also find that leaving out all of the data in epoch $7$ (for the reasons mentioned at the end of Section \ref{sec:performace}) does not significantly change the inference on the lens mass for $\mathcal{M}_{DC}$ -- $M_{\text{L}}=0.54\pm0.06 M_{\odot}$ when all of epoch $7$ is withheld. The mass uncertainty is slightly smaller indicating epoch $7$’s large spread is inflating the uncertainty in the final mass inference. Overall, the lens mass inference is not sensitive to the including or withholding of epoch $7$.

\subsection{Prior sensitivity}\label{sec:prior_sensitivity}

\begin{figure}
    \centering
    \includegraphics[width=\columnwidth]{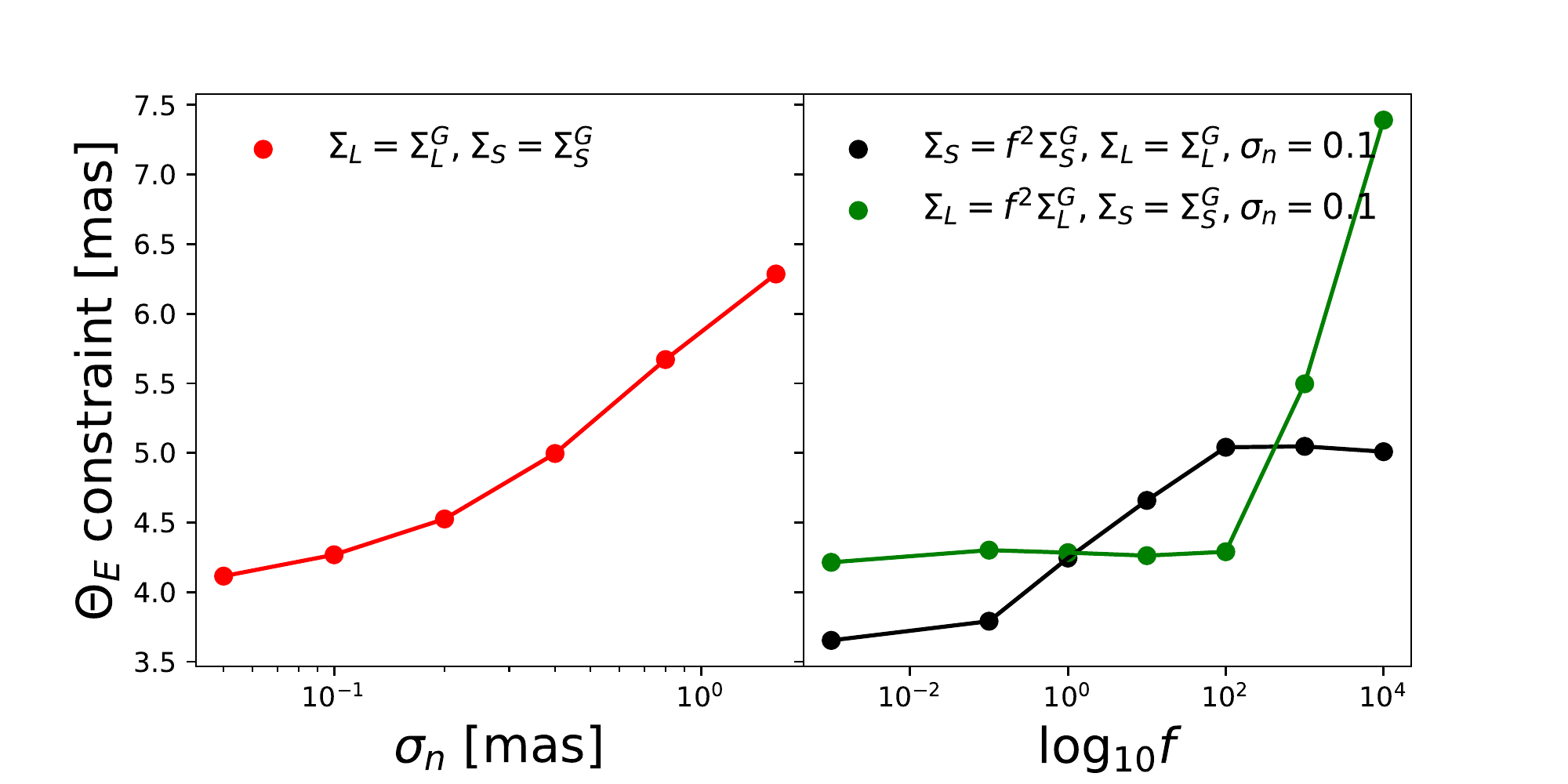}
    \caption[Prior sensitivity]{Sensitivity of the posterior constraint on $\Theta_{\mathrm{E}}$ ($16$th-$84$th posterior percentile) to changes in the priors in the deflection and correlated noise model $\mathcal{M}_{DC}$. \textbf{Left}: Posterior constraint versus $\sigma_{n}$, the standard deviation on the Gaussian prior on each of the correlated noise and white noise parameters. In this case, the priors on the source and lens astrometry are fixed to the GEDR3 $\boldsymbol{\Sigma}^{S}_{G}$ and $\boldsymbol{\Sigma}^{S}_{G}$, respectively. \textbf{Right}: Posterior constraint versus altering the source or lens astrometric Gaussian prior covariance matrix by a multiplicative factor $f$, whilst keeping all other priors fixed.}
    \label{fig:prior_sensitivity}
\end{figure}

We can check how sensitive the posterior inference on $\Theta_{\mathrm{E}}$ is to the prior assumptions in $\mathcal{M}_{DC}$. Specifically, we can test how tightening or relaxing the prior parameter distributions affects the $\Theta_{\mathrm{E}}$ posterior distribution with the view of determining which of our assumptions strongly affect our inferences. For both the lens and source unlensed trajectory, we used informative Gaussian priors from GEDR3. The right panel of Fig. \ref{fig:prior_sensitivity} shows how changing (inflating or shrinking by a multiplicative factor) the GEDR3 prior covariance matrix on the source ($\boldsymbol{\Sigma}^{G}_{S}$) or lens ($\boldsymbol{\Sigma}^{G}_{L}$), changes the $\Theta_{\mathrm{E}}$ posterior constraint. Fig. \ref{fig:prior_sensitivity} shows that the inference on $\Theta_{\mathrm{E}}$ is insensitive to changing $\boldsymbol{\Sigma}^{G}_{L}$. Specifically, the posterior constraint on $\Theta_{\mathrm{E}}$ does not degrade until $\boldsymbol{\Sigma}^{G}_{L}$ is inflated by a factor of $1000^{2}$ (or equivalently multiplying the standard deviation by a factor of $1000$). Moreover, it is also shown that shrinking $\boldsymbol{\Sigma}^{G}_{L}$ (even shrinking by a factor of $1000^{2}$) does not improve the posterior constraint on $\Theta_{\mathrm{E}}$. This means that further data on the lens position, either by future \Gaia{} data releases or further \HST{} monitoring, is unlikely to improve the constraint on $\Theta_{\mathrm{E}}$.

The right panel of Fig.~\ref{fig:prior_sensitivity} also shows the effect of changing the source unlensed trajectory prior covariance matrix $\boldsymbol{\Sigma}^{G}_{S}$, specifically that the outcome is somewhat more sensitive to the prior on the source unlensed trajectory compared with the lens trajectory. Fig.~\ref{fig:prior_sensitivity} shows that inflating the prior by a factor of $10^{2}$ immediately starts to degrade the inference on $\Theta_{\mathrm{E}}$, although the degradation does level off past this point. This is likely because at an inflation level beyond $100^{2}$, the source unlensed trajectory is completely determined by the \HST{} data and the source unlensed trajectory prior becomes uninformative.

More importantly however, shrinking the prior on the unlensed source trajectory does improve the posterior constraint on $\Theta_{\mathrm{E}}$. Specifically, an improvement of a factor of $10^{2}$ in the prior covariance of the source unlensed trajectory would lead to a maximum improvement in the $16$th-$84$th posterior constraint on $\Theta_{\mathrm{E}}$ of $\approx 1$ mas (corresponding to a $7\%$ improvement on $84$th-$16$th percentile constraint on $M_{\text{L}}$). While an improvement of a factor of $10^{2}$ of the source astrometry covariance matrix is unlikely to be achieved, this means that further astrometric data pinning down the unlensing source trajectory either by future \Gaia{} data releases or \HST{} will likely improve the posterior constraint on $\Theta_{\mathrm{E}}$. Assuming a more realistic improvement of $2^{2}$ on the source astrometry covariance matrix, corresponds to a $3\%$ improvement on $16$th-$84$th percentile constraint on $M_{\text{L}}$.

The left panel of Fig.~\ref{fig:prior_sensitivity} shows the effect of inflating or shrinking the standard deviation on the Gaussian prior for the noise parameters ($\sigma_{n}$), both correlated and white. The means of these priors were informed by lens PSF subtraction simulations and WFC3/UVIS instrument precision, in the case of the correlated and white noise terms, respectively. The standard deviations of these priors, ($\sigma_{n}$), were however chosen to be $0.1$ mas. Fig. \ref{fig:prior_sensitivity} shows that the analysis is sensitive to changing $\sigma_{n}$. Specifically, if $\sigma_{n}$ is increased from $0.1$ mas to $0.4$ mas, the $\Theta_{\mathrm{E}}$ $16$th-$84$th posterior percentile degrades by $\approx 1$ mas. In contrast, shrinking $\sigma_{n}$ to $0.01$ mas only marginally improves the $\Theta_{\mathrm{E}}$ $16$th-$84$th posterior percentile, by $\approx 0.2$ mas. This sensitivity is reflected in the posterior distributions of the correlated noise parameters shown in Fig.~\ref{fig:noise_prior_vs_posterior}. This is because for the majority of epochs the size of the correlated noise is not constrained by the data and defaults to the prior distribution, for the adopted value of $\sigma_{n}=0.1$, indicating the priors are informative.

However, for epoch $7$, the data does inform the value of the size of the correlated noise for $\mathcal{M}_{DC}$ and the posterior is slightly shifted to higher values than the prior. This highlights the important balance between picking a reasonably tight prior to inform the size of the correlated noise but one that reflects reasonable uncertainty so the data can alter the value if there is constraining information. The adopted value of $\sigma_{n} = 0.1$ mas in the analysis strikes a reasonable balance between our confidence in the size of the correlated noise at each epoch, whilst also giving sufficient flexibility for the data to further inform the size of the noise. Overall, the posterior constraint on $\Theta_{\mathrm{E}}$ is sensitive to our choice of $\sigma_{n}=0.1$ mas and our prior knowledge on the correlated component of the noise in general. We also investigated changing the boundaries of the flat prior on $\Theta_{E}$ from $20-60$ mas to $0-100$ mas. We found this has no effect on any of the resulting reported parameter posterior percentiles.

\section{The astrophysics of LAWD~37}
\label{sec:mass_implications}

At a distance of $4.6$ pc, LAWD~37 (WD1142-645, 2MASS J11454297-6450297, HIP 57367, Gaia DR3 5332606522595645952, WISEA J114547.32-645033.2), is the second nearest single white dwarf to the sun after van Maanen’s Star. Direct imaging of the region around LAWD~37 with \HST{} shows no evidence of visible companions down to detection limits \citep{Schroeder2000}. Comparing Hipparcos and GDR2 astrometry, \cite{Kervella2019} identified a proper motion anomaly for LAWD~37 which could be explained by a massive companion. However, upon comparison between Hipparcos and the more precise GEDR3 astrometry for LAWD~37, this has largely been ruled out \citep{Lindegren2021}. Overall, at the present time it appears that LAWD~37 has no detectable companions.

Being a close by and bright target (V-band $\approx11.50$), LAWD~37 has been the subject of numerous studies \cite[e.g., ][]{Koester1982,Weidemann1995,Dufour2005,Bergeron2001,Holberg2008, Subasavage2009, Giammichele2012, Sion2009, Subasavage2017, Coutu2019}. As a result, a wealth of photometric and spectroscopic information has been gathered. LAWD~37 has a spectral type DQ indicating a helium dominated atmosphere and the presence of carbon lines in its spectrum. The presence of carbon in DQ white dwarf atmospheres is well explained by models of carbon settling and then being caught up by the helium convection zone bringing it to the surface \citep{Bedard2022}. In line with all helium rich white dwarfs LAWD~37 is expected to have no or trace amounts of hydrogen which is represented by a thin hydrogen layer in models, $q_{H}=10^{-10}$ \citep{Dufour2005}.

\begin{figure}
    \centering
    \includegraphics[width=\columnwidth]{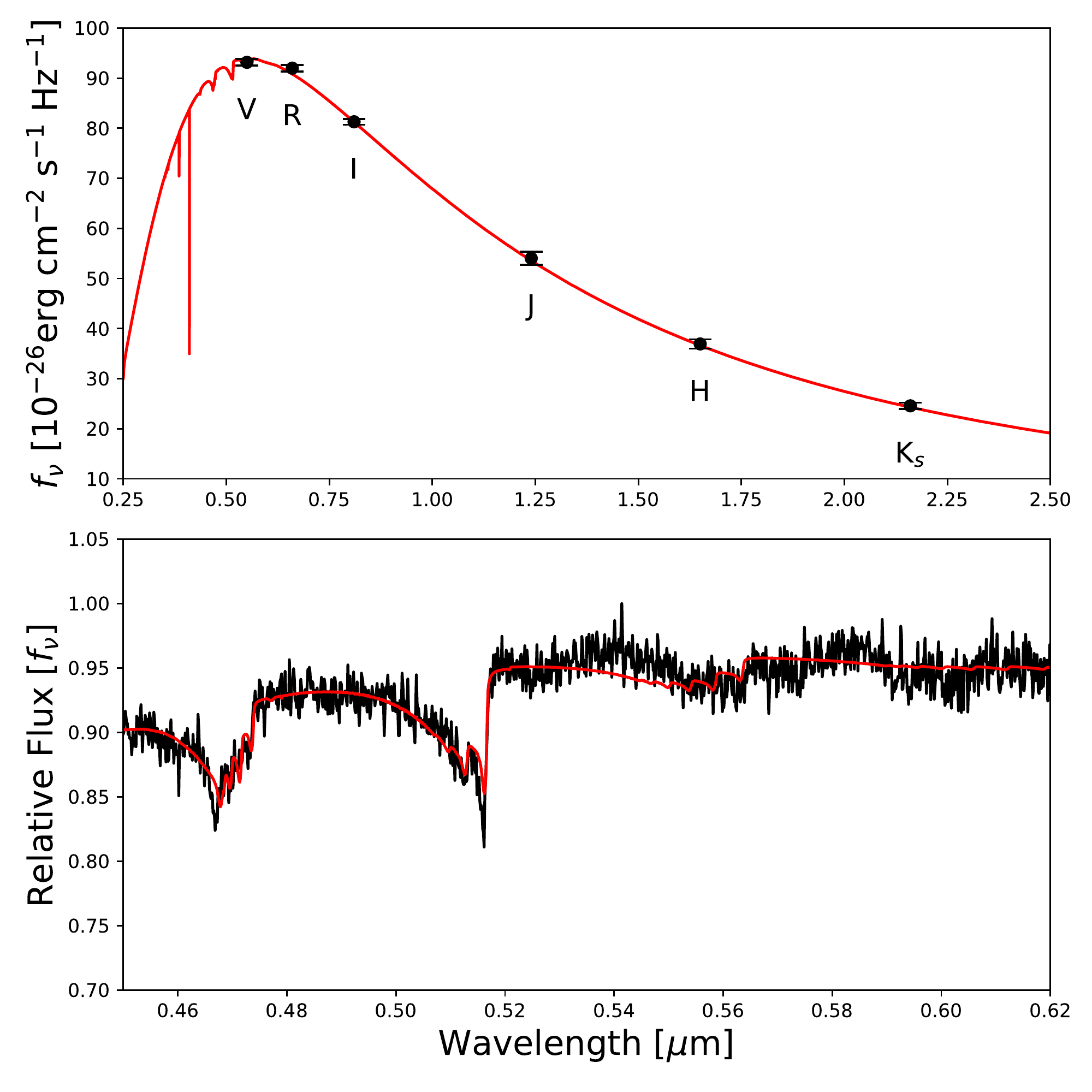}
    \caption{Best fit atmospheric model (red) over the photometric and spectroscopic data (black) of LAWD 37. \protect\textbf{Top}: Photometric fit with VRI bands from \protect\cite{Subasavage2017} and JHK$_{s}$ bands from 2MASS \protect\citep{Skrutski2006}. \protect\textbf{Bottom}: Optical spectrum of LAWD 37 from \protect\citep{Subasavage2017}. The photometry and spectrospcopy of LAWD 37 were fit jointly using the method detailed in Section 4.1 of \protect\cite{Subasavage2017}, and we can see that the models and data are in agreement.}
    \label{fig:spec_phot_fit}
\end{figure}

The atmospheric parameters of LAWD~37 are determined by fitting DQ-type white dwarf atmospheric models \citep{Blouin2018a, Blouin2019} to the broad-band photometry (V,R,I; \citealt{Subasavage2017}, J,H,Ks; 2MASS \citealt{Skrutski2006}), and spectroscopy of LAWD~37. The solid angle and $T_{\rm eff}$ are first adjusted by fitting the model grid to the photometric data assuming an ad hoc photospheric carbon abundance. During this process, the surface gravity (and mass) of the star is calculated using the inferred solid angle of LAWD~37, its distance (from the GEDR3 parallax), and a theoretical mass---radius relationship suitable for CO core white dwarfs with thin hydrogen layers \citep{Bedard2020}. Once a photometric solution is found, the carbon abundance is then adjusted to match the spectroscopy while keeping the other parameters fixed. The photometric fit is then performed once again with this revised carbon abundance, and the whole process (photometric and spectroscopic fits) is repeated until a self-consistent solution is reached. We find $T_{\text{eff}}=7837^{+83}_{-82}$K, $\log g= 7.983\pm{0.016}$, $R=0.0127\pm{0.0003}R_{\odot}$, $M=0.57\pm0.01M_{\odot}$,  and $\log[\text{C/He}]= -5.61\pm0.14$ (see Fig. \ref{fig:spec_phot_fit}). The use of the MRR in this fitting procedure mainly constrains $\log g$ and has a minimal effect on T$_{\text{eff}}$, the solid angle, and therefore the photometric radius determination. This is because the energy distribution is not sensitive to $\log g$.

\begin{figure*}
    \centering
    \includegraphics[width=\columnwidth]{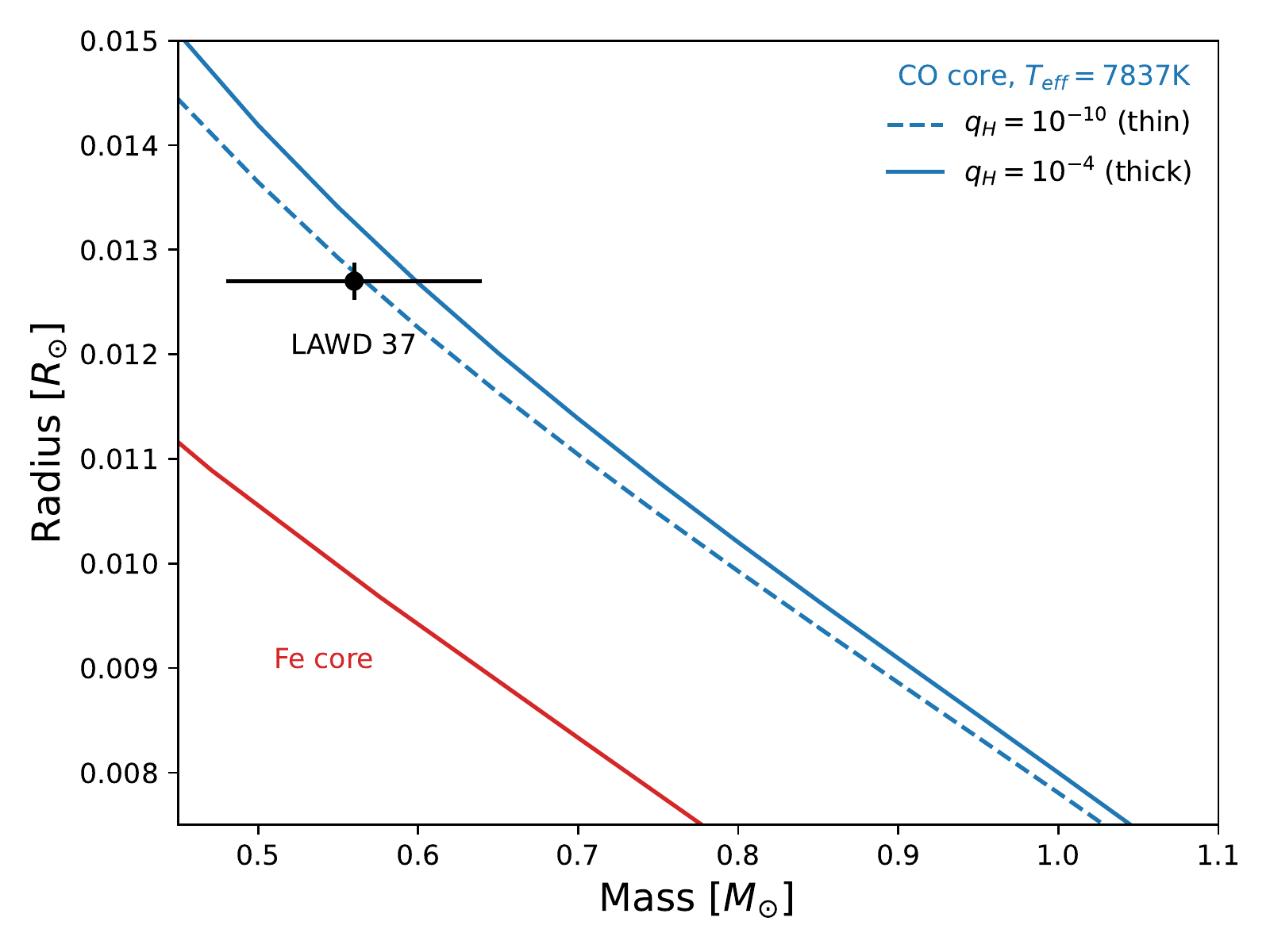}
    \includegraphics[width=\columnwidth]{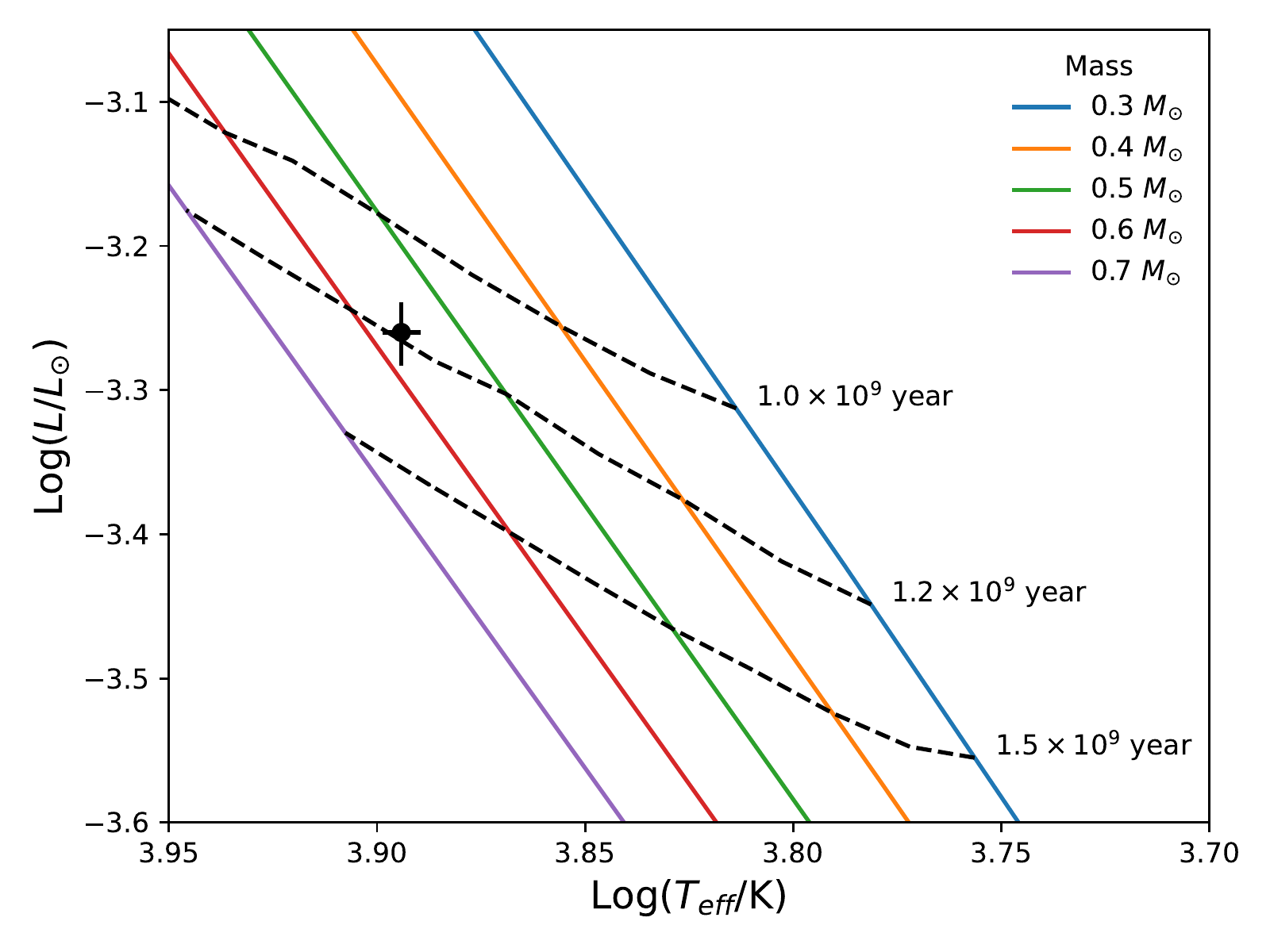}
    \caption[Mass-radius relationship for LAWD 37]{\textbf{Left}: Comparison of the mass of LAWD 37 inferred from the astrometric lensing event with theoretical MRR relationships for white dwarfs. Blue lines show the MRR for CO core white dwarfs with a thin ($q_{H}=10^{-10}$), and thick ($q_{H}=10^{-4}$) hydrogen layer, and effective temperature equal to that of LAWD~37. For comparison, also shown is the theoretical MRR for a zero temperature white dwarf with an iron (Fe) core \citep{Hamada1961}. \textbf{Right}: Hertzsprung-Russell diagram for LAWD~37. Coloured lines show the Montreal cooling tracks for a selection of white dwarf masses. Dashed lines indicate isochrones. The position of LAWD~37 agrees with the inferred microlensing mass of $0.56\pm0.08M_{\odot}$. The implied age of LAWD~37 is $1.15\pm0.04\times10^{9}$ years. }
    \label{fig:lawd_37_MRR}
\end{figure*}

The direct gravitational mass determination from the astrometric microlensing caused by LAWD~37 is completely independent of all atmospheric and evolutionary modeling assumptions. This gravitational mass can therefore be used to test the theoretical models. Fig.~\ref{fig:lawd_37_MRR} shows the position of LAWD~37 on the theoretical MRRs, obtained from the Montreal theoretical cooling tracks\footnote{Obtained from interpolation of the Montreal theoretical cooling tracks available at \url{https://www.astro.umontreal.ca/~bergeron/CoolingModels/}.}, using the gravitational mass from the astrometric microlensing event. Fig.~\ref{fig:lawd_37_MRR} shows excellent agreement between the gravitational microlensing mass from model $\mathcal{M}_{DC}$ and predicted value from the evolutionary theory of CO core white dwarfs, $0.57\pm0.01M_{\odot}$, and assuming a thin hydrogen layer. Also plotted in Fig.~\ref{fig:lawd_37_MRR} is the MRR for CO core white dwarfs with a thick hydrogen layer (not expected for LAWD 37). For comparison, also shown in Fig.~\ref{fig:lawd_37_MRR}, is the theoretical MRR for zero temperature white dwarfs with an iron (Fe) core \citep{Hamada1961}, which is definitively ruled out by the microlensing mass. 

\cite{Coutu2019} observed that the main peak of the mass distribution for DQ white dwarfs in their sample was shifted $
\approx0.05M_{\odot}$ below to that obtained for the DA and DB white dwarfs, suggesting that there could be a problem with the DQ models used in their mass determinations (the same models as those used here). This suspicion was supported by comparing their mass for Procyon B determined with the photometric/spectroscopic method ($0.554M_{\odot}$) with the dynamical mass determination of $0.592M_{\odot}$ determined by \cite{Bond2015}. An alternative explanation for the discrepancy between the mass distributions of DQ white dwarfs and DA/DB stars has recently been offered by \citep[][see their Section 4.3]{Bedard2022} who argued that although all DB stars likely experience carbon dredge-up, this phenomenon is more significant --- and thus more easily detected --- in lower-mass objects, thus explaining the relative deficit of high-mass DQ stars. The good agreement between our gravitational mass for LAWD 37 and the hybrid photometric/spectroscopic value reported here yields support to this interpretation, although the uncertainty on the microlensing mass remains too large to rule out any systematic problem with the current DQ model atmospheres, and this work centers around the analysis of only one object.

Fig.~\ref{fig:lawd_37_MRR} shows the position of LAWD~37 on the theoretical Hertzsprung-Russell diagram (luminosity versus effective temperature) along with theoretical cooling tracks (assuming $q_{H}=10^{-10}$) from the Montreal database for a range of masses. The position of LAWD~37 in the Hertzsprung-Russell diagram is in agreement with the expected position of a white dwarf with the inferred microlensing mass of $0.56\pm0.08 M_{\odot}$.  Also shown in Fig.~\ref{fig:lawd_37_MRR} are isochrones for $1.0$, $1.2$, and $1.5$ $\times10^{9}$years. By interpolation of the Montreal theoretical cooling tracks, the implied cooling age of LAWD~37 is $1.15\pm0.04\times10^{9}$ years.

\section{Discussion and Conclusions}

We have analyzed \HST{} follow-up data of a predicted astrometric microlensing event caused by the nearby white dwarf, LAWD 37. Specifically, we used WFC3/UVIS \HST{} astrometric data of the source in combination with \Gaia{} astrometry (GEDR3) of the source and lens to infer a gravitational mass for LAWD 37 of $0.56\pm0.08 M_{\odot}$. We consider and fit four different models to the data. Models with and without the astrometric deflection term, and then each with and without a correlated noise component due to the lens PSF subtraction within an epoch ($\mathcal{M}_{DC},\mathcal{M}_{DW},\mathcal{M}_{NC},\mathcal{M}_{NW}$). We find the model with the deflection term and correlated noise ($\mathcal{M}_{DC}$) best explains the data according to the overall LOO score.

The model that provides the next best explanation for the data according to the LOO score is the model without an astrometric deflection but with correlated noise ($\mathcal{M}_{NC}$). This model is able to provide an explanation for the deflection signal in the $Y$ direction by increasing the size of the correlated noise above the prior expectation and altering the source proper motion in the $Y$ direction. Therefore, additional follow-up data on the source after the lensing event, which would further constrain the source proper motion, will likely definitively rule out this model. $\mathcal{M}_{NC}$, however, is unable to explain the asymmetric deflection signal in the $X$ direction. This is most prominently seen in epoch $2$ which has a large signal-to-noise deflection in the direction opposite to the source $X$ proper motion direction. 

The model with the deflection and just white noise, $\mathcal{M}_{DW}$, provides a comparatively poor explanation of the data according to the LOO score. $\mathcal{M}_{DW}$ fails to explain the data in epoch $5$ where there is a predicted high amount of correlated noise from the lens PSF subtractions. The failure in epoch $5$ is clear, when compared to $\mathcal{M}_{NC}$, despite $\mathcal{M}_{DW}$ increasing the size of the white noise and epoch $5$ having a large deflection signal. Consequently, because the correlated noise can mimic the deflection signal, $\mathcal{M}_{DW}$ interprets some of the unmodeled correlated noise as additional deflection signal and infers a high mass for LAWD 37 of $M_{L}=0.66^{+0.06}_{-0.05} M_{\odot}$. Overall, we rule out this model, and its mass inference, due to its poor LOO score compared with $\mathcal{M}_{DC}$ and $\mathcal{M}_{NC}$.

We performed checks on the sensitivity of our inferences in our chosen model, $\mathcal{M}_{DC}$, to our prior modeling assumptions. We find that, for $\mathcal{M}_{DC}$, our inference on $M_{\text{L}}$ is not sensitive to the lens GEDR3 prior. This means that collecting additional follow-up astrometric data on the lens (LAWD 37) to improve its astrometric solution is unlikely to improve the inference on $M_{\text{L}}$. In contrast, we find that, for $\mathcal{M}_{DC}$, the inference on $M_{\text{L}}$ is sensitive to the GEDR3 prior on the source astrometry. Therefore, we conclude that collecting additional astrometric data to further constrain the source astrometric solution is likely to improve the inference on $M_{\text{L}}$ by $\approx 3\%$.  For the noise parameters in $\mathcal{M}_{DC}$, both correlated and white, we found our inference on $M_{\text{L}}$ is sensitive to their chosen prior distributions. We found that simultaneously tightening the prior distributions on all the noise parameters does not significantly improve the constraint on $M_{\text{L}}$. 

Overall, we find that the model with the microlensing deflection signal and correlated noise due to the lens PSF subtraction ($\mathcal{M}_{DC}$) provides the best explanation of the data. $\mathcal{M}_{DC}$ provides an inference of the mass of LAWD 37 of $M_{L}=0.56\pm0.08 M_{\odot}$. A no-deflection and correlated noise model, $\mathcal{M}_{NC}$, provides a worse but competitive explanation of the data with $\mathcal{M}_{DC}$. However, $\mathcal{M}_{NC}$, has to increase the size of the correlated noise above our PSF simulation expectations to explain the data in many epochs. Moreover, $\mathcal{M}_{NC}$ cannot explain the high signal-to-noise negative $X$ direction offset in epoch $2$ which is well explained by $\mathcal{M}_{DC}$. We therefore conclude the astrometric deflection signal is present and we measure LAWD 37's mass to be $M_{L}=0.56\pm0.08 M_{\odot}$.

Using the posterior distribution of $\mathcal{M}_{DC}$, we note our results are consistent with the event parameters originally predicted by \cite{McGill2018}. The inferred event parameters (vs the predictions from \citealt{McGill2018}) are: a closes lens-source approach of $396.65\pm0.17$ ($380\pm10$) mas, a time of closest lens-source approach of $2019.86422^{+0.00004}_{-0.00005}$ ($2019.86\pm0.01$) Julian year, and a maximum astrometric shift due to lensing of $2.46\pm0.34$ ($2.8\pm0.1$) mas. The high predicted shift can be attributed to the high photometric mass estimate for LAWD 37 ($0.61M_{\odot}$) used in \cite{McGill2018}. The difference in predicted closest approach can be explained by the poorer quality source astrometry used by \cite{McGill2018}, which was the only source astrometry available at the time.

In conclusion, the gravitational mass for LAWD~37 obtained via astrometric microlensing in this work is in agreement with theoretical MRR and cooling tracks expected from the evolutionary theory of CO core white dwarfs. LAWD~37 is also predicted to cause many more astrometric microlensing events over the coming decades \citep{Bramich2018, Kluter2018b}, which may offer further opportunities to increase the precision of the gravitational mass measurement. This work provides the first ever semi-empirical test of the white dwarf MRR for a single, isolated white dwarf and lends support to current white dwarf evolutionary theory. This work also marks only the third time that the astrometric microlensing effect has been detected via the prediction channel.

This analysis reveals how the followup of future predicted astrometric microlensing events \citep{Bramich2018, Bramich2018b, Kluter2018b,Nielson2018} could be improved. Generally, predicted microlensing events permit targeted and optimised followup campaigns \citep[e.g.,][]{Sahu2017,Zurlo2018,McGill2019b}, because the time of maximum signal can be predicted and hence measurements can be clustered around it. This seems like a sensible approach in the first instance. However, typical predictable microlensing events are caused by nearby bright lenses \citep{Dominik2000}. This means usually we will encounter a scenario similar to this analysis, i.e. where the lens PSF subtraction introduces significant correlated noise into the data. This correlated noise is usually worse at closer lens-source separations (see e.g., Table \ref{tab:corr_noise}), when the deflection signal is largest. Moreover, this analysis shows that the correlated noise can mimic the astrometric deflection signal. Ultimately this meant that epoch $2$ in this analysis, which is in the tails of the event, was critical for measuring the deflection signal, more so than the data at the predicted maximum. The utility of epoch 2 was due to a high signal to noise offset in a direction that could not be explained by alteration of the source's astrometric parameters. 

With the power of hindsight, the importance of taking observations at epoch 2's time could have been determined ahead of planning the \HST{} observations. This is because all of the lensing systems parameters (lens and source astrometry and lens mass estimate) were already known, and the size of the correlated noise introduced by the lens PSF subtraction could be roughly estimated ahead of time. However, at the time of closest approach, LAWD 37 was at the edge of the “sun avoidance zone”, which caused some increased noise, and also a shortened visibility period during an orbit. The small range of available ORIENTs at this configuration made it impossible to keep the diffraction spike too far from the source. This turned out to be the most important factor in elevating the noise, and there was no way to avoid this. This aspect of the observation planning is also extremely difficult to simulate in advance.  

In principle, an observation strategy with shorter intervals between epochs, and with fewer exposures per epoch could be used. This would provide more constraints on the source astrometry at a greater range of epochs and hence a degree of redundancy of epoch in the case that residuals prove to be irremovable. However, in the case of followup campaigns carried out with \textit{HST}, including the followup of the event presented in this paper, one orbit is the minimum that can be used per visit. A future line of research could work out if \HST{} \citep[or perhaps even the \textit{James Webb Space Telescope};][]{Gardner2006} observation times which optimally constrain the lens mass and/or rule out purely correlated noise explanations of the data can be determined ahead of time. This could be tested on future predicted microlensing events.

\section*{Acknowledgements}

We would like to thank the two referees for their detailed reports that both improved the clarity of the paper. Based in part on observations made with the NASA\slash ESA {\it Hubble Space Telescope}, obtained at STScI, which is operated by the Association of Universities for Research in Astronomy, Inc., under NASA contract NAS~5-26555. Support for this research was provided by NASA through grants from STScI\null. {\it HST} data used in this paper are available from the Mikulski Archive for Space Telescopes at STScI (\url{https://archive.stsci.edu/hst/search.php}) under proposal IDs 15705, 15961 and 16251. We thank Howard Bond and Ed Nelan for their help at various stages of the project. We thank Daniel Bramich, Tom Brown, Ed Nelan, Łukasz Wyrzykowski, and Paul Hewett for useful comments that improved the clarity of the paper. PM acknowledges studentship funding support from the United Kingdom Science and Technology Facilities Council (STFC) and the Cambridge Centre for Doctoral Training in Data Intensive Science (CDT-DIS). MBN acknowledges support from the UK Space Agency. This research is based on observations made with the NASA/ESA \textit{Hubble Space Telescope} obtained from the Space Telescope Science Institute, which is operated by the Association of Universities for Research in Astronomy, Inc., under NASA contract NAS 5–26555. This work presents results from the European Space Agency (ESA) space mission \Gaia{}. \Gaia{} data are being processed by the \Gaia{} Data Processing and Analysis Consortium (DPAC). Funding for the DPAC is provided by national institutions, in particular the institutions participating in the \Gaia{} Multi-Lateral Agreement (MLA). The \Gaia{} mission website is \url{https://www.cosmos.esa.int/gaia}. The \Gaia{} Archive website is \url{http://archives.esac.esa.int/gaia}.

%%%%%%%%%%%%%%%%%%%%%%%%%%%%%%%%%%%%%%%%%%%%%%%%%%
\section*{Data Availability}

The data underlying this article will be shared on reasonable request to the corresponding author.

%%%%%%%%%%%%%%%%%%%% REFERENCES %%%%%%%%%%%%%%%%%%

\bibliographystyle{mnras}
\bibliography{Bibliography}

%%%%%%%%%%%%%%%%%%%%%%%%%%%%%%%%%%%%%%%%%%%%%%%%%%

%%%%%%%%%%%%%%%%% APPENDICES %%%%%%%%%%%%%%%%%%%%%

\appendix

\section{Leave-One-Out cross-validation approximation}\label{sec:loo}

In this Section we explain how the LOO PSIS approximation is applied to the models considered in this work. This section closely follows \cite{Vehtari2017} and \cite{Burkner2020}. The goal of the approximation, given that we have obtained $S$ posterior samples from fitting $\mathcal{M}_{TN}$ to the full data set $\mathcal{D}$, $\{\boldsymbol{\theta}^{s}\}_{s=1}^{S}$, we wish to approximate,
\begin{equation}
    p(D_{i}|\mathcal{D}_{-i},\mathcal{M}_{TN})= \int p(D_{i}|\boldsymbol{\theta},\mathcal{D}_{-i},\mathcal{M}_{TN})p(\boldsymbol{\theta}|\mathcal{D}_{-i},\mathcal{M}_{TN})d\boldsymbol{\theta}.
    \label{eq:target}
\end{equation}
Specifically, we want to use $\{\boldsymbol{\theta}^{s}\}_{s=1}^{S}$ to approximately evaluate Eq. (\ref{eq:target}) instead of having to refit the model (which is expensive) with the $i$th data point left-out to obtain $\{\boldsymbol{\theta}^{s}_{-i}\}_{s=1}^{S}$. This is achieved by using an importance sampling approximation and re-weighting $\{\boldsymbol{\theta}^{s}\}_{s=1}^{S}$ accordingly. 

Importance sampling is a Monte Carlo method used to compute expectations. We would like to evaluate the expectation,
\begin{equation}
    \mathbb{E}_{f}[h(\theta)] = \int h(\theta)f(\theta)d\theta,
\end{equation}
where $f$ is some hard-to-sample-from distribution. Instead of using samples from $f$ we can use samples from an easier-to-sample-from proposal distribution $g$, $\{\theta^{s}_{g}\}^{S}_{s=1}$, provided we know the the ratio between $f$ and $g$ which is $r(\theta^{s}_{g})=f(\theta^{s}_{g})/g(\theta^{s}_{g})$. The importance sampling approximation is then,
\begin{equation}
    \mathbb{E}_{f}[h(\theta)] \approx \frac{\sum^{S}_{s=1}r(\theta^{s}_{g})h(\theta^{s}_{g})}{\sum^{S}_{s=1}r(\theta^{s}_{g})}.
    \label{eq:importance_sampling}
\end{equation}
The quality of this approximation is sensitive to the distribution of the importance weights, $r(\theta^{s}_{g})$. If the proposal distribution, $g$, is not representative of the target distribution $f$, then the importance weights become unstable. This instability comes from the importance weights being dominated by a few extreme values which leads to them having a large or infinite variance. Ultimately this leads to a poor importance sampling approximation \citep{Vehtari2015}. To mitigate this problem, $r(\theta^{s}_{g})$ are fitted to a Pareto distribution and the extreme values are removed and are replaced with draws from the fitted Pareto distribution, $\tilde{r}(\theta^{s}_{g})$. These smoothed weights, $\tilde{r}(\theta^{s}_{g})$, are dropped in as a replacement for $r(\theta^{s}_{g})$ in Eq. (\ref{eq:importance_sampling}). This is called Pareto Smoothed Importance Sampling \citep[PSIS;][]{Vehtari2015}. In addition to stabilising the importance sample approximation, PSIS also provides a diagnostic on the quality of the importance sampling approximation. This diagnostic is the fitted shape parameter, k, of the Pareto distribution. k traces the number of finite moments of the importance weights distribution and therefore the quality of the PSIS approximation. \cite{Vehtari2015} find that $k<0.7$ indicates PSIS will work well whereas a value of $k>0.7$ indicates the PSIS approximation is likely to be poor and should not be used.

For approximating the LOO score, our hard-to-sample-from target distribution is $p(\boldsymbol{\theta}|\mathcal{D}_{-i},\mathcal{M}_{TN})$. This is the posterior distribution obtained when fitting the model to the data set with the $i$th data point left-out. This is hard to sample from because it is expensive and we would have to refit the model, which we want to avoid. h is $p(D_{i}|\boldsymbol{\theta},\mathcal{D}_{-i},\mathcal{M}_{TN})$.  Our easy to sample from proposal distribution is $p(\boldsymbol{\theta}|\mathcal{D},\mathcal{M}_{TN})$ which we readily have samples from - $\{\boldsymbol{\theta}^{s}\}_{s=1}^{S}$. The only remaining task is to compute the ratio between the target and proposal distributions.
\begin{align}
    r(\boldsymbol{\theta}) &= \frac{p(\boldsymbol{\theta}|\mathcal{D}_{-i},\mathcal{M}_{TN})}{p(\boldsymbol{\theta}|\mathcal{D},\mathcal{M}_{TN})}\propto  \frac{p(\boldsymbol{\theta}|\mathcal{M}_{TN})p(\mathcal{D}_{-i}|\boldsymbol{\theta},\mathcal{M}_{TN})}{p(\boldsymbol{\theta}|\mathcal{M}_{TN})p(\mathcal{D}| \boldsymbol{\theta},\mathcal{M}_{TN})}\\&= \frac{p(\mathcal{D}_{-i}|\boldsymbol{\theta},\mathcal{M}_{TN})}{p(\mathcal{D} | \boldsymbol{\theta},\mathcal{M}_{TN})} .
    \label{eq:posterior_ratios}
\end{align}
Here we have used Bayes rule which states that the ratio of the posteriors is proportional to the ratio of the prior $\times$ likelihood, and the priors terms have cancelled. For all models the likelihood factorises over the different epochs (Eq. \ref{eq:likelihood}). If the left-out data point, $D_{i}$, is the $j$th data point in epoch $\tilde{e}$, all other epochs cancel from the likelihood ratio in Eq. (\ref{eq:posterior_ratios}) and,
\begin{equation}
    r(\boldsymbol{\theta}) \propto\frac{p(\mathcal{D}_{-i}|\boldsymbol{\theta},\mathcal{M}_{TN})}{p(\mathcal{D}| \boldsymbol{\theta},\mathcal{M}_{TN})} = \frac{1}{p(D_{\tilde{e},j}|D_{\tilde{e},-j}, \boldsymbol{\theta},\mathcal{M}_{TN})}.
\end{equation}
The ratio is proportional to the inverse of the  likelihood of the left-out data point \citep{Burkner2020}. In all models considered in this work, $p(D_{\tilde{e},j}|D_{\tilde{e},-j}, \boldsymbol{\theta},\mathcal{M})$ is the product of two multivariate Gaussian distributions (in the $X$ and $Y$ directions) as described in Eq. (\ref{eq:like_direction}). Using the results from  \cite{Sundararajan2001}, \cite{Burkner2020} and Eq. (\ref{eq:like_direction}), this likelihood can be computed efficiently as, 
\begin{align}
    \begin{split}
    \log p(D_{\tilde{e},j}|D_{\tilde{e},-j}, \boldsymbol{\theta},\mathcal{M}_{TN}) &= -\log 2\pi \tilde{\sigma}_{\tilde{e},j} \\ &- \frac{1}{2\tilde{\sigma}_{\tilde{e},j}}\Bigg(\left(\left[\boldsymbol{X}_{\tilde{e}}\right]_{j}-\tilde{X}_{\tilde{e},j}\right)^{2} \\ &+ \left(\left[\boldsymbol{Y}_{\tilde{e}}\right]_{j}-\tilde{Y}_{\tilde{e},j}\right)^{2}\Bigg).
    \end{split}
\end{align}
Here, $\left[\bullet\right]_{j}$ is the $j$th component of vector $\bullet$. The conditional means, $\tilde{X}_{\tilde{e},j}$ and $\tilde{Y}_{\tilde{e}, j}$, and conditional standard deviation $\tilde{\sigma}_{\tilde{e},j}$ are given by \cite{Burkner2020} as,

\begin{figure}
    \centering
    \includegraphics[width=\columnwidth]{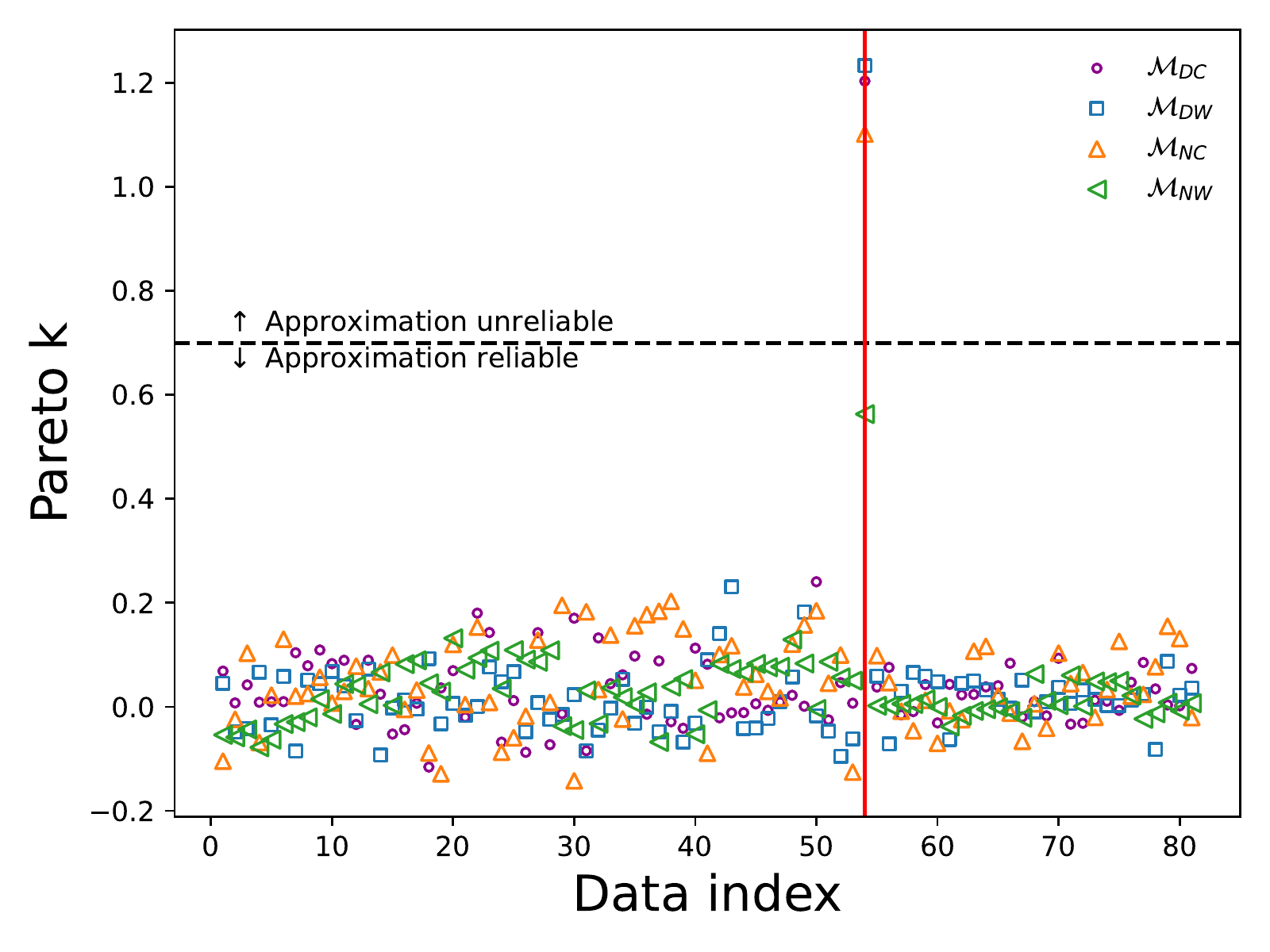}
    \caption[PSIS Pareto k values]{Pareto k values in the PSIS-LOO approximation used for each model. Different colours indicate different models. Different sized makers are used so points on top of each other can be differentiated. Dashed horizontal line indicates the threshold value of Pareto k ($0.7$) used to judge if the importance sampling approximation is reliable. For all models, the importance sampling approximation fails for the data point with index $54$ indicated as a vertical red line. To obtain the LOO score for data point $54$ the model was refit.}
    \label{fig:pareto_k}
\end{figure}

\begin{equation}
    \tilde{X}_{\tilde{e},j} =\left[\boldsymbol{X}_{\tilde{e}}\right]_{j} - \frac{\left[(\boldsymbol{\Sigma}^{N}_{\tilde{e}})^{-1}(\boldsymbol{X}_{\tilde{e}}-\boldsymbol{X}^{T}_{\tilde{e}})\right]_{j}}{\left[(\boldsymbol{\Sigma}^{N}_{\tilde{e}})^{-1}\right]_{jj}},
\end{equation}
\begin{equation}
\tilde{Y}_{\tilde{e},j} = \left[\boldsymbol{Y}_{\tilde{e}}\right]_{j} - \frac{\left[(\boldsymbol{\Sigma}^{N}_{\tilde{e}})^{-1}(\boldsymbol{Y}_{\tilde{e}}-\boldsymbol{Y}^{T}_{\tilde{e}})\right]_{j}}{\left[(\boldsymbol{\Sigma}^{N}_{\tilde{e}})^{-1}\right]_{jj}},
\end{equation}
and,
\begin{equation}
    \tilde{\sigma}_{\tilde{e},j} = \frac{1}{\left[(\boldsymbol{\Sigma}^{N}_{\tilde{e}})^{-1}\right]_{jj}}.
\end{equation}

Here, we have dropped the dependence of $\boldsymbol{\Sigma}^{N}_{\tilde{e}}$, $\boldsymbol{X}^{T}_{\tilde{e}}$ and $\boldsymbol{Y}^{T}_{\tilde{e}}$ on $\boldsymbol{\theta}$ for brevity. $\left[\bullet\right]_{jj}$ denotes the $j$ diagonal element of the matrix $\bullet$.

Using these weights with PSIS, we can then approximate Eq. (\ref{eq:target}) for each left-out data point using the posterior samples from the full dataset, $\{\boldsymbol{\theta}^{s}\}_{s=1}^{S}$. For each left-out data point, we obtain the Pareto $k$ diagnostic. If for any left-out data point $k>0.7$, we do not use the PSIS approximation, and instead perform a full refit to evaluate Eq. (\ref{eq:target}) exactly. Fig. \ref{fig:pareto_k} shows the Pareto $k$ diagnostic for each left-out data point for all models. For every model the approximation was reliable for all left-out data points apart from $D_{53}$. For all models we therefore computed this LOO term exactly by refitting the model with $D_{53}$ left-out. Overall, with PSIS, we obtained an approximate LOO score for each model with only one additional refit of the model compared to the $81$ refits required to compute LOO exactly. 

\begin{figure}
    \centering
    \includegraphics[width=\columnwidth]{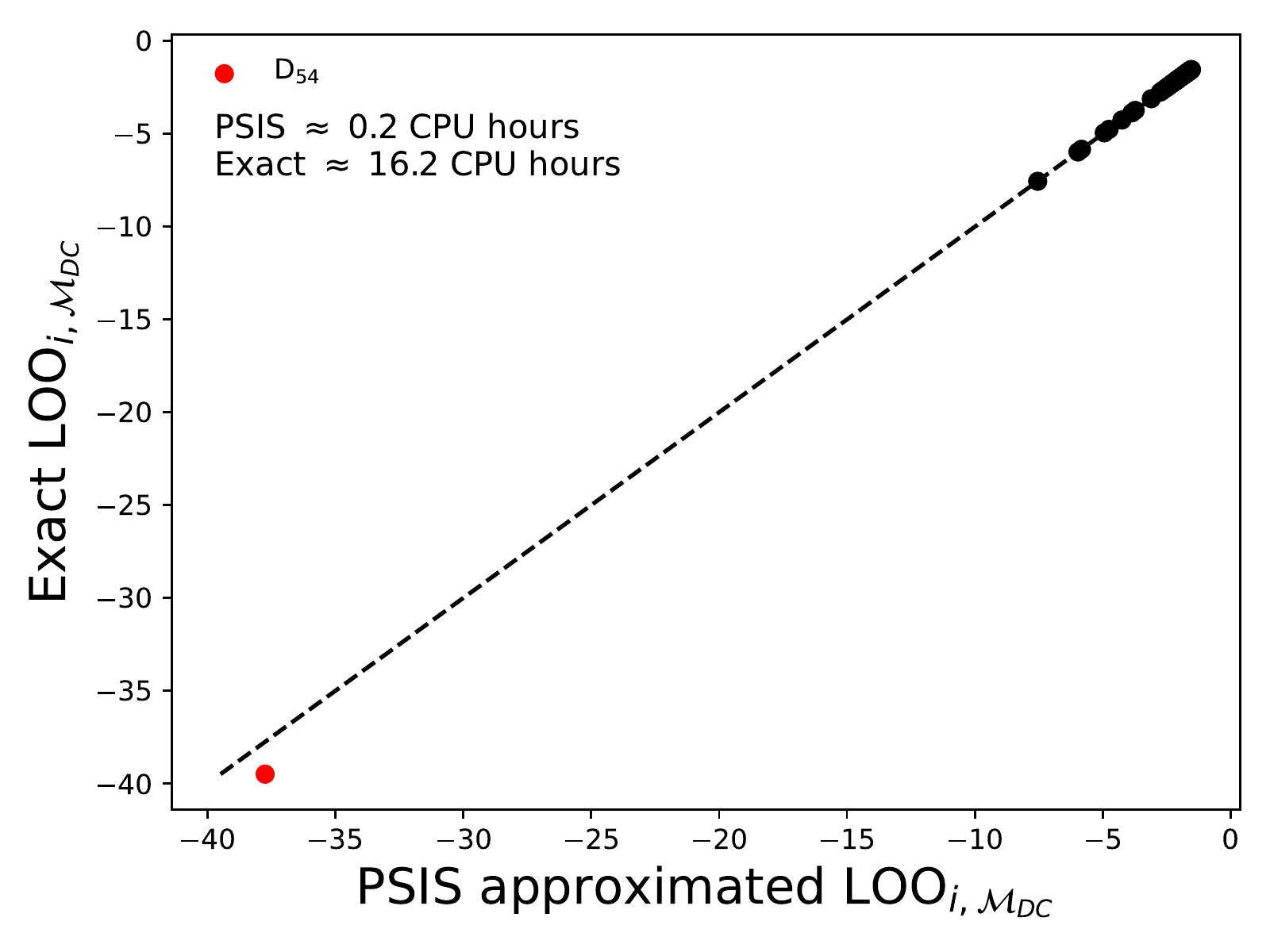}
    \caption[PSIS approximation accuracy]{PSIS approximation versus exact computation of the LOO terms for $\mathcal{M}_{DC}$. The PSIS is good agreement with the exact values. The dashed line is the one-to-one relationship. The data point that failed the PSIS approximation $D_{53}$, is marked in red. The approximate CPU computation times for the exact and PSIS methods are shown as text.}
    \label{fig:PSIS_vs_exact}
\end{figure}

To further check the safety of the PSIS approximation, we computed all LOO terms for the model, $\mathcal{M}_{DC}$, exactly and compared them to the PSIS approximation. Fig. \ref{fig:PSIS_vs_exact} shows the comparison of the PSIS approximation and the exactly computed LOO terms from $\mathcal{M}_{DC}$. The PSIS approximation matches the exact LOO computation well for all left-out data points apart from $D_{54}$. Encouragingly, $D_{54}$ is the same data point that was flagged by PSIS as the approximation being unreliable (Fig. \ref{fig:pareto_k}). Consequently, the model was refitted and the exact LOO value was used for $D_{54}$.  Fig \ref{fig:PSIS_vs_exact} also shows that the PSIS approximation was $\approx 80$ times faster than the exact LOO computation.

The large negative value of $\text{LOO}_{54,\mathcal{M}_{DC}}$ indicates that $D_{54}$ is an outlying data point which is not well predicted by $\mathcal{M}_{DW}$. The failure of PSIS approximation for $D_{54}$ means that the removal of this data point changes the posterior by a large amount compared with the removal of all other data points. This is unsurprising as $D_{54}$ belongs to epoch $7$ where the source was close to a charge bleed column (Fig. \ref{fig:lawd_74_mosaic}), and the estimated size of the correlated noise is high (Table \ref{tab:corr_noise}), due to the specialized fitting procedure used in this epoch (see Section \ref{sec:hst_data}). $D_{54}$ can also be seen to be the furthest outlying brown pentagon marker in Fig. \ref{fig:posterior_over_data}. Other than $D_{54}$'s membership to epoch 7, there is nothing else in the reduction that explains its significant outlying position. One possible explanation could be that if a cosmic ray landed at the very center of the source PSF when $D_{54}$ was taken, it could have corrupted the source position and would be very hard to identify.

\section{Consistency with \Gaia{}}\label{sec:consitency_with_gaia}

\begin{figure*}
    \centering
    \includegraphics[width=\textwidth]{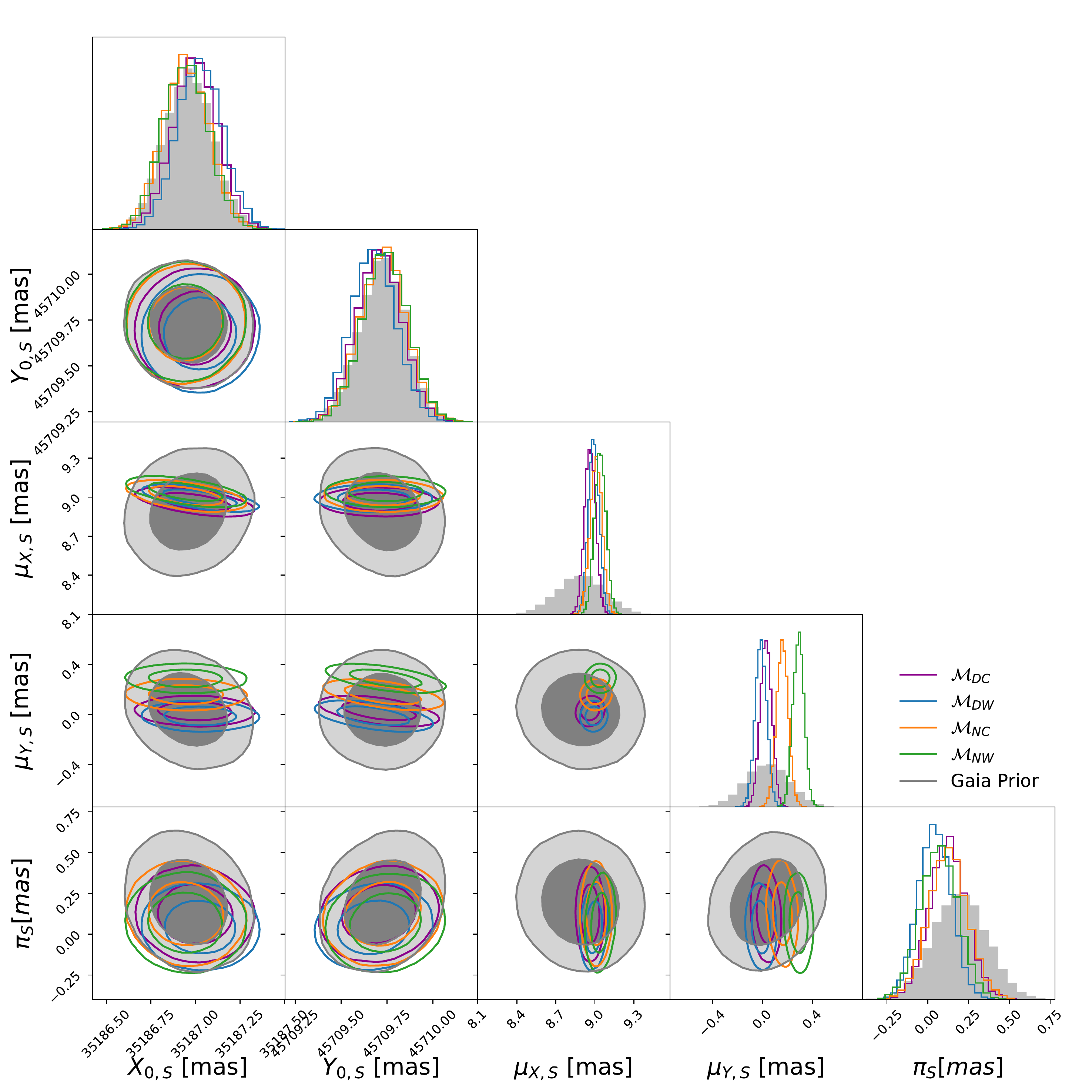}
    \caption[\Gaia{} source prior versus posterior]{GEDR3 priors versus posterior inferences on all source astrometric parameters for all the models considered. The plot is a full version of the densities shown in Fig. \ref{fig:source_prior_vs_posterior}.}
    \label{fig:full_source_ast}
\end{figure*}

\begin{figure*}
    \centering
    \includegraphics[width=\textwidth]{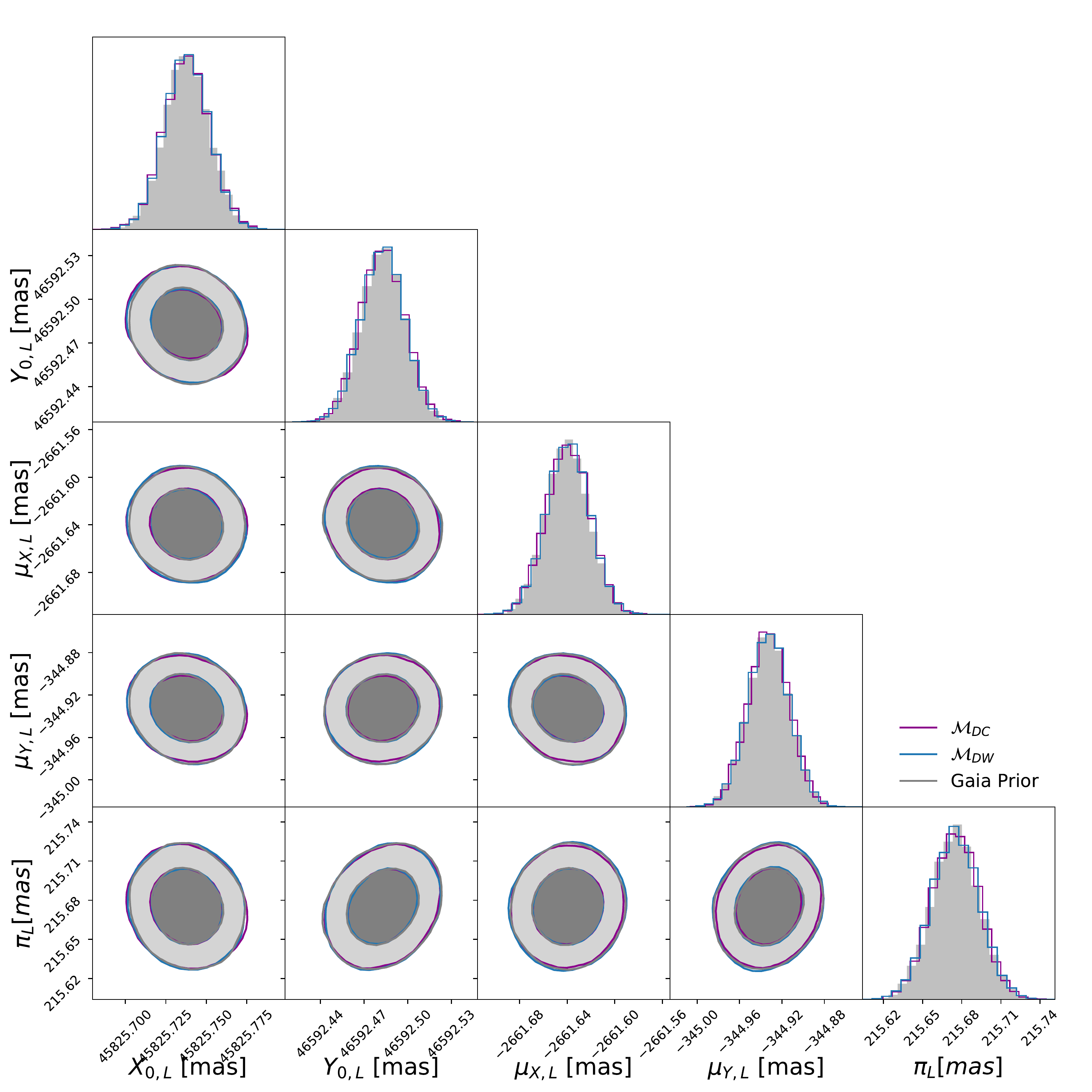}
    \caption[\Gaia{} lens prior versus posterior]{GEDR3 priors versus posterior inferences on all the lens (LAWD 37) astrometric parameters for the two deflection models which contain the astrometry as free parameters. For both these models, the \HST{} data provides no further constraint from GEDR3.}
    \label{fig:full_lens_ast}
\end{figure*}

This Section contains the full corner plots for both the source (Fig. \ref{fig:full_source_ast}) and lens (Fig. \ref{fig:full_lens_ast}) prior and posterior distributions of their astrometric parameters. For both the lens and source, and all models, it can be seen that the posteriors are in good agreement with the GEDR3 priors.

\bsp
\label{lastpage}
\end{document}